\documentclass[twocolumn,fleqn]{aastex701}
\usepackage{amsmath}
\usepackage{hyperref}
\newcommand{\angstrom}{\text{\normalfont\AA}}

\shorttitle{Broad \ion{Mg}{2} Galaxies}
\shortauthors{LaMassa et al.}

\begin{document}

\title{Are Broad \ion{Mg}{2} Galaxies Fading AGN?}

\correspondingauthor{Stephanie LaMassa}

\author[0000-0002-5907-3330]{Stephanie M. LaMassa}
\affiliation{Space Telescope Science Institute
3700 San Martin Dr.,
Baltimore, MD, 21218, USA}
\email{slamassa@stsci.edu}

\author[0000-0002-3719-940X]{Michael Eracleous}
\affiliation{Department of Astronomy and Astrophysics and
Institute for Gravitation and the Cosmos, Penn State University,
525 Davey Lab, 251 Pollock Road, University Park, PA 16802, USA}
\email{mxe17@psu.edu}

\author[0000-0001-8557-2822]{Jessie Runnoe}
\affiliation{Department of Physics \& Astronomy, Vanderbilt
  University,
  2301 Vanderbilt Place, Nashville, TN 37235, USA;
  Department of Life and Physical Sciences, Fisk University,
  1000 17th Avenue N, Nashville, TN 37208, USA}
\email{jessie.c.runnoe@vanderbilt.edu}

\author[0000-0002-3683-7297]{Benny Trakhtenbrot}
\affiliation{School of Physics and Astronomy,
 Tel Aviv University, 
 Tel Aviv 69978, Israel}
\email{bennyt@tauex.tau.ac.il}

\author[0000-0003-3242-7052]{Gisella de Rosa}
\affiliation{Space Telescope Science Institute
3700 San Martin Dr.,
Baltimore, MD, 21218, USA}
\email{gderosa@stsci.edu}

\begin{abstract}
We analyzed a sample of 494 galaxies from the Sloan Digital Sky
Survey (SDSS) I-IV that have unusual properties. They display broadened
\ion{Mg}{2}\,$\lambda$2798 line emission, as would be expected of Type 1 active galactic
nuclei (AGN). However, they have
weak or absent broad Balmer lines though broadened Balmer lines often accompany
broadened ultraviolet lines in Type 1 AGN spectra. We performed a spectral
decomposition of these ``broad \ion{Mg}{2} galaxies'' to assess
whether they can be considered fading AGN. The Balmer lines are more
responsive to changes in the AGN continuum than the \ion{Mg}{2}
line. As the AGN power source fades, broad H$\beta$ and
then H$\alpha$ will weaken and eventually disappear, leaving broad \ion{Mg}{2} as a signpost of
a more active past. In support of this interpretation, we found
that broad \ion{Mg}{2} galaxies have weaker AGN luminosities and lower
accretion rates (parameterized by the Eddington ratio) than
typical Type 1 quasars from SDSS. The Eddington ratio of broad
\ion{Mg}{2} galaxies are similar to dim-state changing-look AGN, with
an average value $\lesssim$1\%, which is the threshold at which
the inner region of a geometrically thin accretion disk is thought to evaporate and be replaced by a
geometrically thick radiatively inefficient accretion
flow in stellar-mass black holes. Broad \ion{Mg}{2} galaxies may
represent the penultimate stage in a changing-look AGN sequence prior
to the complete disappearance of all broad lines.
\end{abstract}


\section{Introduction}

At the center of nearly every massive galaxy lies a supermassive black
hole that can be millions to billions of times more massive than the
Sun. These black holes are observed as Active Galactic Nuclei (AGN)
when they grow via a steady supply of fuel from an accretion
disk. According to the \citet{shakurasunyaev1973} model, the accretion
disk powering AGN is optically thick and geometrically thin though
other accretion models, like geometrically thick radiatively
inefficient accretion flows \citep{shapiro1976} and advection-dominated accretion flows \citep{narayanyi1994}, may
describe the power source of sub-Eddington AGN
\citep{lasota1996,quataert1999,narayan2005,narayanmcclintock2008}.

Galaxies harboring growing supermassive black holes can be identified
via spectroscopic signatures. Gas near the
vicinity of the accretion disk orbits rapidly, producing broadened
emission lines in rest-frame ultraviolet to near-infrared spectra. AGN
where this ``broad line region'' is visible are categorized as Type 1
AGN. In some systems, broad lines are not present but the existence of
an accreting black hole can be inferred by prominent narrow emission
lines whose ratios are consistent with
photoionization by the continuum from an accretion disk \citep[using diagnostic ``BPT
diagrams'';][]{bpt,Kewley2001,Kewley2006,Kauffmann2003}. These galaxies are classified as Type 2 AGN.

The theory of AGN unification provides a framework for linking Type 1
and Type 2 AGN by positing that both are powered by the same central
engine (an accreting supermassive black hole) but that a circumnuclear, parsec
scale torus of dust and gas obscures the accretion disk and broad line region
from view when the AGN is oriented edge-on to the line of sight \citep{antonucci1993,urry1995,netzer2015}. While this model is broadly
successful in describing observed properties of AGN, and
polarization measurements of some Type 2 AGN reveal hidden
broad lines in polarized light consistent with this model's predictions
\citep[e.g.,][]{antonucci1985,miller1991,tran1992,tran2001,tran2003,zakamska2005},
this description is not complete.

A class of objects dubbed ``changing-look AGN'' exhibit remarkable
variability on the timescale of years where broad emission lines,
typically Balmer lines, and the non-stellar continuum strengthen or fade \citep[e.g.,][]{tohline1976,denney2014,lamassa2015,runnoe2016,MacLeod2016,MacLeod2019,yang2018,Green2022,ricci2023,Wang2024,Zeltyn2024,Guo2025,Dong2025}. Most of this emission
variability can be ascribed to changes in the power of the central
engine to support the production of broad lines rather than
variable obscuration. AGN spectral type can thus be a product of both obscuration
and AGN evolution, with changing-look AGN providing insight into
accretion physics and the AGN duty cycle.

There are a class of galaxies discovered in the Sloan Digital Sky
Survey \citep[SDSS;][]{York2000, gunn06} that have broad \ion{Mg}{2} emission at 2798\angstrom\ in the
ultraviolet but narrow Balmer lines or weak or no broad emission lines
(\citealp{roig2014}; see also \citealp{lamassa2016a}). Such
broadened \ion{Mg}{2} emission would be expected in AGN, but the broad
Balmer lines associated with Type 1 AGN are lacking. We propose that
these ``broad \ion{Mg}{2} galaxies'' are signposts of fading AGN, on
the tail end of changing-look AGN spectral variability. According to
theoretical models for the Broad Line Region, \ion{Mg}{2} exhibits lower
responsivity than the Balmer lines \citep{KoristaGoad2004}, so we may expect a broad remnant
to \ion{Mg}{2} as the broad Balmer lines disappear \citep[see the
changing-look AGN sequence described in][]{Guo2020a}. There is also some observational
evidence for this hypothesis: examples of changing-look AGN observed
at sufficiently high redshift show that as the broad Balmer liness fade, a broad component to
\ion{Mg}{2} remains
\citep[e.g.,][]{MacLeod2016,MacLeod2019,Graham2020,Green2022,Zeltyn2024,zeltyn2026},
though changing-look AGN with variable broad \ion{Mg}{2} emission have
been discovered \citep{guo2019,guo2024}.

In this paper, we test the hypothesis that broad \ion{Mg}{2} galaxies
are fading AGN. We select a sample of broad \ion{Mg}{2} galaxies from
SDSS and compare their properties with a control sample of Type 1
quasars selected from SDSS in the same redshift range, as well as with a
sample of SDSS changing-look AGN. We investigate the distribution of
the AGN physical parameters in these samples to assess where broad
\ion{Mg}{2} galaxies lie on the AGN continuum. We qualitatively
compare the properties of broad \ion{Mg}{2} galaxies, SDSS quasars,
and changing-look AGN with theoretical predictions. Throughout, we
assume a flat $\Lambda$CDM cosmology where H$_{\rm 0}$ = 70 km/s/Mpc
and $\Omega_{\rm m}$ = 0.3.

\section{Sample Selection}

\subsection{Defining a broad \ion{Mg}{2} galaxy sample}
We selected broad \ion{Mg}{2} galaxy candidates 
from SDSS Legacy \citep{York2000, gunn06} through SDSS-IV Data Release
16 \citep{blanton17, ahumada20} by querying CASJobs \citep{li08}
for sources classified as
``Galaxies'' by the SDSS spectroscopic pipeline and lying in the
redshift range $0.35 < z < 0.75$, which captures both the \ion{Mg}{2}
and H$\beta$ spectral regions in observed frame optical
spectra. This selection garnered 1,413,876 galaxies.

We performed simple fitting of these spectra to identify
candidates for a subsequent full spectral decomposition with \textsc{PyQSOFit} \citep{guo18,shen19}
which is a more time intensive process. We shifted all spectra to the rest
frame and calculated a residual spectrum by subtracting the fitted
SDSS spectral model
from the observed spectrum. The SDSS ``Galaxy'' model, which is a
linear template model derived from eigenspectra produced by principal-component analysis
of SDSS galaxy
training samples \citep{bolton2012}, does not
include broad emission lines so any broad \ion{Mg}{2} emission would
become apparent in this residual spectrum. We then
calculated the signal-to-noise ratio (S/N)
in the spectrum around the \ion{Mg}{2} line, from 2775 - 2825
\angstrom\ using the Python \textsc{specutils} routine
\textsc{snr}. For sources where the S/N in the spectrum
around the \ion{Mg}{2} line exceeded 0.5, we fitted the \ion{Mg}{2}
complex in the residual spectra with a double Gaussian model.  We retained sources where the
full-width at half maximum (FWHM) of the broadest Gaussian component
exceeded 2000 km s$^{-1}$.

This initial pass through the SDSS spectra provided 4150 spectra
  of broad
\ion{Mg}{2} galaxy candidates, including multiple
  observations of some galaxies that we later vet. We fitted these
  spectra with
\textsc{PyQSOFit} on
\textsc{SciServer} \citep{Taghizadeh-Popp20}, as described in detail in Section
\ref{fit_pyqsofit}.  To ensure consistency in the measurements of
spectral properties we analyze in this work, we select comparison SDSS quasar
and changing-look AGN samples whose spectra we model with
\textsc{PyQSOFit} using the same version of the software and input
parameters that we use for the broad \ion{Mg}{2} galaxies.

\subsection{Comparison SDSS Type 1 Quasar Sample}
To compare the properties of broad \ion{Mg}{2} galaxies with typical
Type 1 quasars, we selected a control sample of quasars from the
Sixteenth Data Release of the SDSS Quasar Catalog \citep{lyke2020} for
analysis (hereafter ``SDSS quasars''). From this catalog of over 750,000 quasars, we selected 5,000
as a baseline for comparison with the broad
\ion{Mg}{2} galaxy sample. Since about half of the \ion{Mg}{2} 
galaxies are at a redshift between $0.35 < z < 0.6$ and the other half
are between $0.6 < z < 0.75$, we randomly selected 2500 SDSS quasars
from each of these redshift bins that we then fit with
\textsc{PyQSOFit} as described below. We note that after
  applying the quality control cuts to the spectral fits that are
  discussed below, we retain a comparable fraction of quasars and
  broad \ion{Mg}{2} galaxies in the
  low and high redshift bins, with
  44\% (48\%) of quasars (broad \ion{Mg}{2} galaxies)  below $z < 0.6$
  and 56\% (52\%) above $z > 0.6$.

\subsection{Comparison Changing-Look AGN Sample}
To assess whether broad \ion{Mg}{2} galaxies are on the tail end of
the spectrum of AGN variability, we compare their properties with
those of changing-look AGN. We selected a comparison sample of
changing-look AGN from SDSS so that the spectral resolution and data
quality are consistent with the broad \ion{Mg}{2} galaxy spectra. We
therefore require that the changing-look AGN used for our comparison sample have SDSS spectra in both
the bright and dim state. We select our comparison sample from SDSS
I-IV, matching the data releases from which we selected our broad
\ion{Mg}{2} galaxy sample, so changing-look AGN discovered from SDSS V \citep[e.g.,][]{Zeltyn2024}
are not included in this analysis.

We searched the literature for changing-look AGN from SDSS I-IV between a
redshift of 
$0.35 < z < 0.75$ and identified 17 AGN for analysis from the
following sources:
\citet{MacLeod2016,MacLeod2019,Wang2018,Graham2020,Green2022}. We fetched the original spectra
from SDSS and fitted them with \textsc{PyQSOFit} as detailed in
Section \ref{fit_pyqsofit}.

We
only retain the spectral fit parameters for further analysis if the
broad \ion{Mg}{2} line is securely detected so that we match the
sample selection of the changing-look AGN comparison sample with that
of the broad \ion{Mg}{2} galaxy sample, as described
in detail below. Of the 17 changing-look
AGN we fitted, we obtained an acceptable fit for 16 of them in either
the bright or dim state; we did not obtain a secure broad \ion{Mg}{2}
line detection for AGN J100220.17+450927.3 from \citet{MacLeod2016}
due to the \ion{Mg}{2} line being too close to the blue end of the spectrum
(see Section \ref{control_cut})
and thus do not include this source in our analysis.
We list this comparison sample of 16 changing-look AGN, along with associated
literature references, in Table \ref{cl_agn_sample}. We bold-face the
spectra that passed our quality control checks and are thus used in
our subsequent analysis.

 In principle, the dim-state changing-look AGN could also be independently
  selected as broad \ion{Mg}{2} galaxies since these dim-state AGN could be expected
  to have weak or absent broad Balmer emission and be host galaxy
  dominated. However, all changing-look AGN that we considered for
  this analysis are classified by the SDSS spectroscopic pipeline as
  ``QSO''s in both the bright and dim state due to prominent broad
  H$\beta$ or H$\alpha$ emission that lingers even after the AGN
  continuum fades and the broad Balmer lines weaken. Thus they are not
  included in the list of SDSS spectroscopic ``Galaxies'' that we
  queried to identify broad \ion{Mg}{2} galaxy candidates. On the
  other hand, the spectra of changing-look AGN could be included in
  the SDSS quasar comparison sample. By design, we chose to consider
  the changing-look AGN spectra separately to assess how the
  properties of broad \ion{Mg}{2} galaxies compares with bright-state
  and dim-state changing-look AGN, with the overall SDSS quasar
  population serving as a control sample. While there are no duplicate
spectra between our SDSS quasar sample and changing-look AGN sample,
the SDSS quasar sample includes spectra of two changing-look AGN
observed at other epochs.

\begin{deluxetable*}{lllllll}
\tablecaption{\label{cl_agn_sample}Changing-Look AGN Sample\tablenotemark{1}}
\tablehead{\colhead{Source} & \colhead{$z$} & \colhead{Plate} &
 \colhead{MJD} & \colhead{FiberID}  & \colhead{State} &
 \colhead{Reference}} 
\startdata
J002311.06+003517.5 & 0.422 & 4219 & 55480 & 852 & bright &
\citet{MacLeod2016,Green2022} \\
& & \bf{7864} & \bf{56979} & \bf{983} & \bf{dim} & \\
J020514.77-045639.7 & 0.363 & \bf{4393} & \bf{55944} & \bf{358} & \bf{bright} &
\citet{Green2022} \\
& & 9327	& 57723 & 246 & dim \\
J022556.07+003026.7 & 0.504 & \bf{3615} & \bf{55445} & \bf{617} & \bf{dim} &
\citet{MacLeod2016} \\
& & \bf{3615}	& \bf{56544} & \bf{669} & \bf{bright} \\
J025410.09+034912.5 & 0.774 & \bf{4256} & \bf{55477} & \bf{278} &
\bf{dim} &
\citet{Graham2020}\\
& & \bf{9415}	& \bf{58029} & \bf{117} & \bf{bright} \\
J025619.03-004501.0 & 0.722	& \bf{410} & \bf{51877} & \bf{83} & \bf{bright} &
\citet{Graham2020} \\
& & \bf{9372}	& \bf{58074} & \bf{236} & \bf{dim} \\
J080138.67+423355.2 & 0.771 & \bf{3683} & \bf{55178} & \bf{268} & \bf{bright} & 
\citet{Graham2020} \\
& & \bf{8277}	& \bf{57073} & \bf{242} & \bf{dim} \\
J081632.13+404804.6 & 0.701 & \bf{760} & \bf{52264} & \bf{216} & \bf{bright} &
\citet{Graham2020} \\
& & \bf{8293}	& \bf{57361} & \bf{393} & \bf{dim} \\
J095750.03+530106.0 & 0.437 & \bf{903} & \bf{52400} & \bf{383} & \bf{dim} &
\citet{Graham2020} \\
& & \bf{8182}	& \bf{56993} & \bf{843} & \bf{bright} \\
J100302.62+193251.3 & 0.467 & \bf{2371} & \bf{53762} & \bf{620} & \bf{dim} &
\citet{Green2022} \\
& & \bf{9578} & \bf{57817} & \bf{290} & \bf{bright} \\
J105513.88+242553.7 & 0.496 & \bf{2357} & \bf{53793} & \bf{128} &
\bf{bright} &
\citet{MacLeod2019,Green2022} \\
& & \bf{10467} & \bf{58138} & \bf{724} & \bf{dim} \\
J113111.15+373709.4 & 0.448 & \bf{2036} & \bf{53446} & \bf{227} & \bf{bright} &
\citet{Graham2020} \\
& & \bf{8874}	& \bf{57426} & \bf{801} & \bf{dim} \\
J141324.28+530527.0 & 0.456 & 1325 & 52762 & 133 & dim & \citet{Wang2018}\\
&  & \bf{7339} & \bf{57518} & \bf{735} & \bf{bright} & \\
J214613.31+000930.8 & 0.621 & 1107 & 52968 & 358	& dim &
\citet{MacLeod2016} \\
& & \bf{4196}	& \bf{55478} & \bf{774} & \bf{bright} \\
J225240.37+010958.7 & 0.534 & 676 & 52174 & 442 & dim &
\citet{MacLeod2016}\\
& & \bf{4294}	& \bf{55500} & \bf{45} & \bf{bright} \\
J233317.38-002303.4 & 0.513 & 681 & 52199 & 114 & dim &
\citet{MacLeod2016} \\
& & \bf{4212}	& \bf{55447} & \bf{312} & \bf{bright} \\
J234623.42+010918.1 & 0.509 & \bf{683} & \bf{52524} & \bf{457} & \bf{dim} &
\citet{Green2022} \\
& & \bf{9178}	& \bf{58081} & \bf{477} & \bf{bright} \\
\enddata
\tablenotetext{1}{Changing-look AGN from the literature with SDSS spectra
 in the dim and bright states that we fitted with
 \textsc{PyQSOFit}. If the fitted parameters passed the quality
 control checks described in the main text, the spectral fit
 parameters are used in the analysis for this paper. These spectra are highlighted with
 bold-faced text. ``Plate'', ``MJD'', and ``FiberID'' refer to the
 SDSS spectral identifiers, i.e., the identification numbers of the
 spectroscopic plate and fiber and the Modified Julian Date of the
 observation.}
\end{deluxetable*}

\section{Analysis}

\subsection{Fitting SDSS Spectra with PyQSOFit}\label{fit_pyqsofit}
\textsc{PyQSOFit} \citep{guo18,shen19} is a \textsc{python} package
that decomposes the spectrum of an AGN into quasar continuum and host
galaxy components and fits Gaussian models to individual emission line components, both
narrow and broad, that are
defined by the user. The quasar continuum itself is made up of several
possible components: power-law continuum, a polynomial component, and a Balmer
Continuum \citep{Dietrich2002}. Ultraviolet \ion{Fe}{2} \citep{vestergaard01,salviander07,tsuzuki06} and optical
\ion{Fe}{2} templates \citep{boroson92}, which form a
pseudo-continuum, are also included in the fitting.  We used all these quasar components
except for the
Balmer Continuum because we found that adding this component did not
improve our spectral fits.

\begin{figure*}
\includegraphics[scale=0.45]{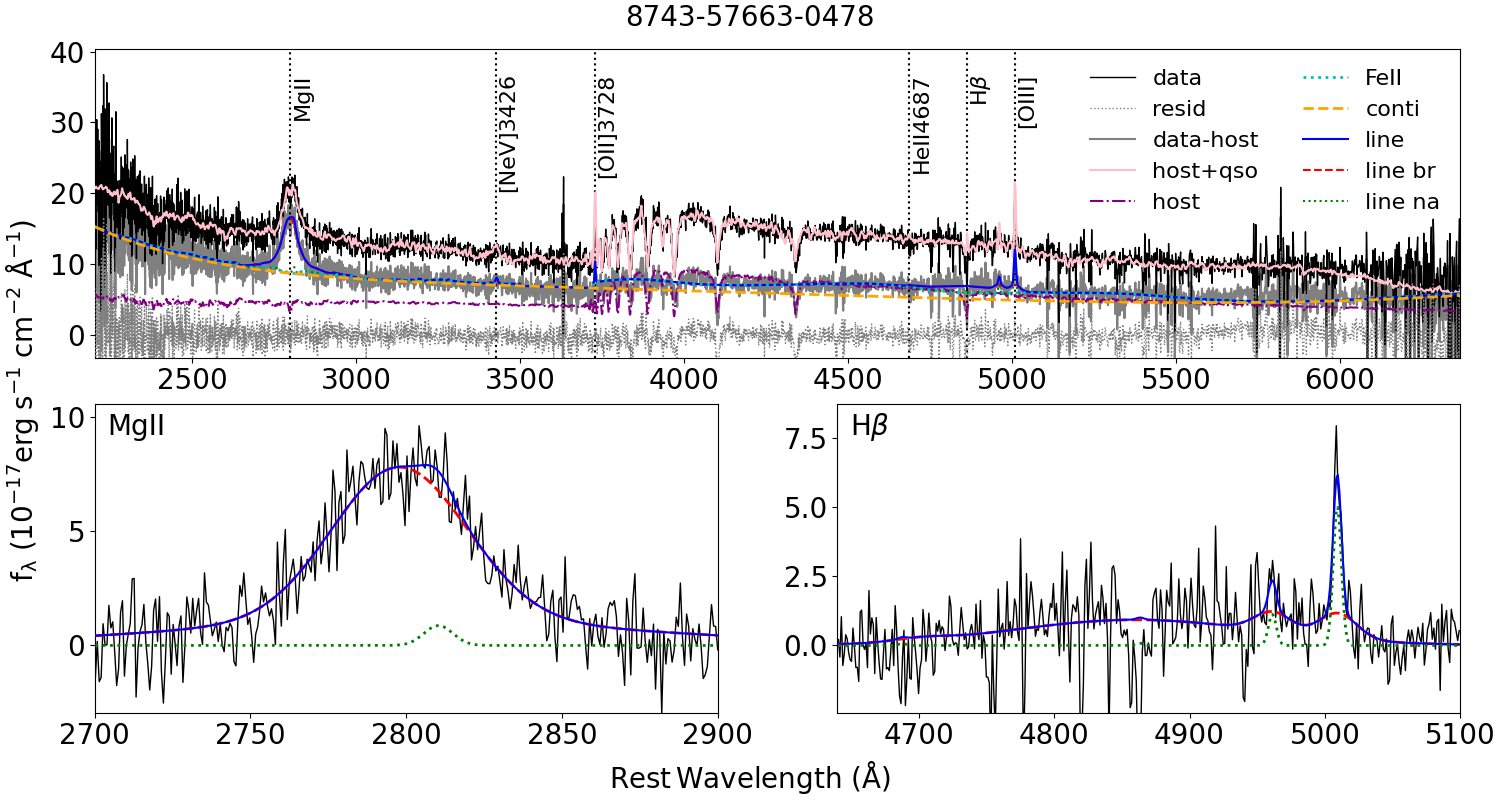}\\
\includegraphics[scale=0.45]{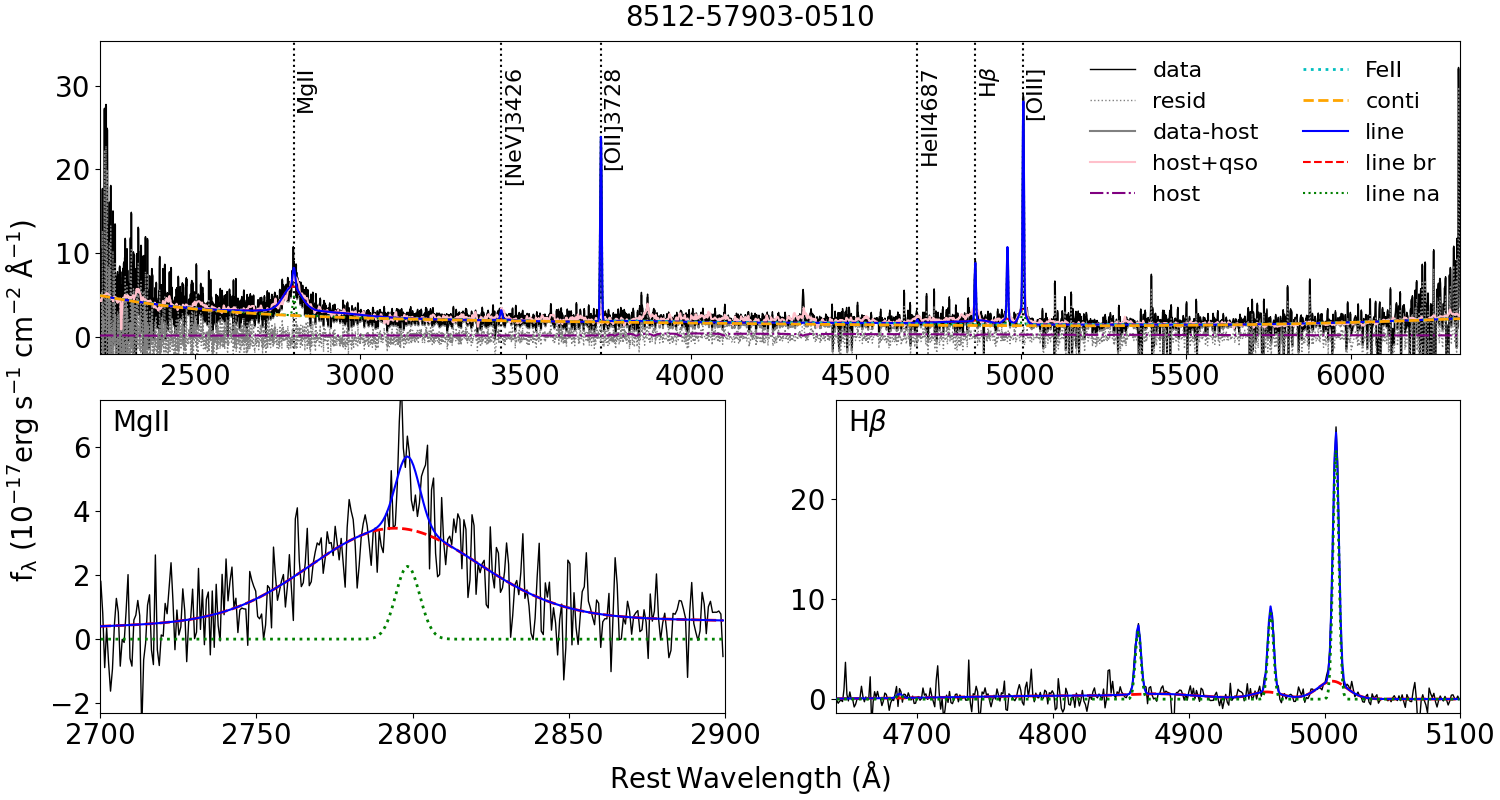}
\caption{\label{pyqsofit_examples} Examples of
  broad \ion{Mg}{2} galaxies fitted with \textsc{PyQSOFit}
  \citep{guo18,shen19} using E-MILES
  spectral templates \citep{vazdekis16} for the host galaxy component. This method
  reliably decomposes the host (purple, dot-dashed line) and AGN
  continuum (orange dashed line and
  cyan dotted line) emission for a range of galaxy types, including
  post-starburst galaxies (top plot) and emission line galaxies
  (bottom plot). The emission lines that \textsc{PyQSOFit} fits, if
  present in the spectra, are indicated in the top panels.  Note that neither the [\ion{Ne}{5}]\,$\lambda$3426 nor
  the \ion{He}{2}\,$\lambda$4687 narrow emission lines were detected in either galaxy and
  their transitions are marked for illustrative purposes only. The zoom-in plots highlight the \ion{Mg}{2} and
  H$\beta$ complex emission line fits where both broad (red dashed) and
  narrow (green dotted) line fits are indicated. Note that
  fitted emission lines need to pass quality control checks as
  described in the main text in order for the fit parameters to be
  included in the catalog. For instance, in the top
  galaxy, \textsc{PyQSOFit} accommodates a slight asymmetry in the
  \ion{Mg}{2} profile with a weak, narrow Gaussian component, but this
  \ion{Mg}{2} narrow line is not significant and is rejected by the
  quality control cuts discussed in the main text. However, a narrow
  component to \ion{Mg}{2} is significantly detected in the bottom galaxy. The spectra are
named by their SDSS observing identifiers: plate number - MJD - fiber
id (see Table \ref{cat_descr}).}
 \end{figure*}

The default method for fitting the host galaxy in \textsc{PyQSOFit} is
to use principal component analysis with an input set of 
eigenspectra based on SDSS galaxy spectra \citep{yip04a} and SDSS
quasar spectra \citep{yip04b}. However, \textsc{PyQSOFit} has the
option to incorporate user-provided galaxy spectral templates for the
host galaxy decomposition. We used this option to input 53 E-MILES \citep{vazdekis16} spectral templates ranging in ages from 30 Myr
- 14 Gyr, assuming BaSTI isochrones \citep{pietrinferni2004,pietrinferni2006}, a Kroupa revised broken powerlaw Initial Mass
Function \citep{kroupa2001}, and metallacity [M/H]=+0.06.\footnote{We downloaded
  E-MILES model spectra with these input parameters from the following
  website:
  \url{https://research.iac.es/proyecto/miles/pages/webtools/tune-ssp-models.php}}
We found
that the E-MILES templates, along with the \textsc{PyQSOFit}-provided
quasar eigenspectra, did an acceptable job fitting the range of host
galaxy properties in our broad \ion{Mg}{2} galaxy sample, including
post-starburst galaxies and emission line galaxies (see Figure \ref{pyqsofit_examples} for
examples).

We defined the following emission line complexes for \textsc{PyQSOFit}
to fit if present in the spectra, where the reported transitions
are vacuum wavelengths:
the \ion{Mg}{2}\,$\lambda$2780 unresolved doublet, the
  [\ion{Ne}{5}]\,$\lambda$3426
line, the [\ion{O}{2}]\,$\lambda$3728 unresolved doublet, and the H$\beta$
complex, including \ion{H}{1}\,$\lambda$4862, \ion{He}{2}\,$\lambda$4687 and
[\ion{O}{3}]\,$\lambda\lambda$5008,4960 (with separate components for the line
core emission and potential broadened emission due to outflows); and
the H$\alpha$ complex, including \ion{H}{1}\,$\lambda6564$, and [\ion{N}{2}]\,$\lambda\lambda$6549,6585. We set
up \textsc{PyQSOFit} to use up to two Gaussian components to model any
broad \ion{Mg}{2}, H$\beta$ and H$\alpha$ line emission. The fitted line
center is a free parameter that is allowed to vary from the nominal
line center within a tolerance of 0.5 - 1.5\%, depending on whether
the line is narrow or broad. \textsc{PyQSOFit} fits both broad and
narrow emission lines at the input transitions, where we used the default FWHM value of 1200 km s$^{-1}$ to differentiate between
broad and narrow emission lines. This default dividing line is based on the results from \citet{hao2005} that
show this value as an empirical demarcation between Type 1 and Type 2 AGN
in SDSS.

We used Monte Carlo (MC) sampling with 50 trials in \textsc{PyQSOFit} to calculate the
measurement errors, which is standard practice when fitting
  hundreds to thousands of optical spectra from SDSS and other
  ground-based surveys with \textsc{PyQSOFit}, where the number of MC
  samples used in the literature range from 25 \citep[e.g.,][]{wu2022,Zeltyn2024,aydar2026}
  to 50 \citep[e.g.,][]{timlin2020,rakshit2020,guo2020b,jin2023}. In
  the Appendix, we report the results of our investigation to test the
  impact of increasing the number of MC samples on the
  reported uncertainties, measurement precision, and measurement
  accuracy. We find that 50 MC samples strikes a reasonable balance
  between compute time and measurement accuracy as increasing the
  number of MC trials does not generally significantly improve results
and incurs a significant cost in computing time.

\subsection{Selection Criteria for Broad \ion{Mg}{2} Detection}\label{control_cut}
We applied a series of quality cuts on the fit results to
retain sources where a broad \ion{Mg}{2} line was securely detected
and well measured. We require that all \ion{Mg}{2} broad line fit parameters
(central wavelength, line $\sigma$, FWHM, equivalent width, flux)
are non-zero and exceed the error on the parameter, and that the
parameter error is non-zero.

We also ensure that the spectral
decomposition near the \ion{Mg}{2} line is well measured by applying a cut on the measured AGN continuum luminosity at 3000 \angstrom\ ($\lambda L_{\rm
  3000}$)  and by checking
that the spectral window to fit \ion{Mg}{2} is sufficiently wide. We
retain sources where $\lambda L_{\rm
  3000}$ and its associated error is greater than zero and where $\lambda L_{\rm
  3000}$ exceeds the error on $\lambda L_{\rm 3000}$.\footnote{Since
\textsc{PyQSOFit} reports AGN continuum luminosities and errors in log
space, we retain sources where the error on $\lambda L_{\rm 3000}$ is
less than 0.3 dex.} To check that the \ion{Mg}{2} line is not cut off at the
blue end of the spectrum, we require that the fitted line center $-3\sigma$ is greater than the minimum wavelength of the spectrum.

Finally, we search for well-fitted \ion{Mg}{2} lines by requiring that
the reduced $\chi^2$ value of the fit within the \ion{Mg}{2} line complex be
under 1.5. For a broad \ion{Mg}{2} line to be considered ``detected,''
we impose a \ion{Mg}{2} S/N $\geq$1.5 cut, where the line S/N reported
by \textsc{PyQSOFit} refers to the
amplitude of the Gaussian fit to the line divided by the noise in the residual of the line
fit (i.e., data - model).

We apply the same cuts to the other emission lines fitted by
\textsc{PyQSOFit}, except for the reduced $\chi^2$ criterion since the
reduced $\chi^2$ is computed for the entire line complex, where some
individual lines are well measured and others are not,
making it an unreliable indicator of fit quality for other emission
lines. Here we check that the H$\beta$ and H$\alpha$ line
complexes (if present) are not cut off at the red end of the spectrum: the fitted
line center $+3\sigma$ has to be less than the maximum wavelength of
the spectrum. We only include emission line measurements that meet the
quality control criteria in the catalogs and the analysis below.

\subsection{Multiple Spectra of Broad \ion{Mg}{2} Galaxies}
A number of broad \ion{Mg}{2} galaxies had spectra from multiple
observations. By design, our comparison SDSS quasar sample and
dim-state and bright-state changing-look AGN samples do not have repeat
observations of the same source since we chose representative spectra
of changing-look AGN in each state.

When applying the spectral quality cuts to the \textsc{PyQSOFit}
results of the broad \ion{Mg}{2} galaxy candidates, we were
left with 520 spectra of 494 unique sources. Twelve galaxies had
multiple spectral epochs that were fit independently. We checked the
fit parameters and spectral decomposition of these sources by eye and
found that the fit parameters were consistent among these spectra
given the quality of the spectra (i.e., there was no evidence
of large variability between epochs). We then chose one spectrum to use as
the representative spectrum for the galaxy which was the
spectrum where the \ion{Mg}{2} S/N was the highest or, in one case, where there were
smaller residuals around the fit to H$\beta$.

\section{Results}
For a source to be included in our broad \ion{Mg}{2} galaxy sample, 
comparison SDSS quasar sample and comparison changing-look AGN samples, we
require a strong broad \ion{Mg}{2} line detection and reliable \ion{Mg}{2} 
line measurements. These quality cuts garner 494 unique broad
\ion{Mg}{2} galaxies, 3613
SDSS quasars, 15 bright-state
changing-look AGN, and 11 dim-state changing look AGN.

\subsection{PyQSOFit Spectral Decomposition Results}

We release four catalogs that summarize the properties of the samples
analyzed in this work: broad \ion{Mg}{2} galaxies, SDSS Type 1
quasars, SDSS changing-look AGN in the bright state, and SDSS
changing-look AGN in the dim state. These catalogs include the AGN continuum luminosities at 3000
\angstrom\ and 5100 \angstrom\ and the line fit parameters for \ion{Mg}{2} and
the following species (if detected according to our quality control
cuts described above): [\ion{Ne}{5}], [\ion{O}{2}],
\ion{He}{2}, H$\beta$, [\ion{O}{3}], H$\alpha$, [\ion{N}{2}]. We list
a description of the catalog columns in Table \ref{cat_descr}.

\begin{deluxetable*}{p{0.25\linewidth}  p{0.6\linewidth}}
\tablecaption{\label{cat_descr}Catalog Description\tablenotemark{1,2}}
\tablehead{\colhead{Column Name} & \colhead{Description}} 
\startdata
Name & Source name is defined by SDSS observation identifiers: Plate - MJD - FiberID. \\
RA & Right Ascension (degrees). \\
Dec & Declination (degrees). \\
plateid & SDSS spectrcopic plate number of observation. \\
mjd & Modified Julian Date of observation. \\
fiberid & SDSS spectroscopic fiber identification number. \\
redshift & SDSS spectroscopic redshift. \\
L3000, L3000\_err & AGN continuum luminosity at 3000 \angstrom\ 
($\lambda L_{\rm 3000}$) and associated error. Units are erg s$^{-1}$
and are reported in log space. \\
L5100, L5100\_err & AGN continuum luminosity at 5100 \angstrom\ 
($\lambda L_{\rm 5100}$) and associated error. Units are erg s$^{-1}$
and are reported in log space. \\
{\it line}\_FWHM, {\it line}\_FWHM\_err & Full-width-half-maximum of
Gaussian fit to emission line and associated error. Units are km s$^{-1}$. \\
{\it line}\_sigma, {\it line}\_sigma\_err & Sigma of
Gaussian fit to emission line and associated error. Units are km s$^{-1}$. \\
{\it line}\_EW, {\it line}\_EW\_err & Equivalent width of
Gaussian fit to emission line and associated error, reported in rest frame. Units are
\angstrom. \\
{\it line}\_wavecen, {\it line}\_wavecen\_err & Fitted center of
emission line and associated error. Units are
\angstrom. \\
{\it line}\_flux, {\it line}\_flux\_err & Flux of emission line. Units are
10$^{-17}$ erg s$^{-1}$ cm$^{-2}$. \\
{\it line}\_SN, {\it line}\_SN\_err & S/N of emission line, calculated
by dividing the Gaussian amplitude by the noise in the residual
of the model fit. \\
MBH, MBH\_err & Black hole mass ($M_{\rm BH}$)  calculated with the broad component of
the \ion{Mg}{2} line and $\lambda L_{\rm 3000}$ using the
formula from \citet{Shen2024}. Units are $M_{\rm \odot}$. \\
L\_Edd, L\_Edd\_err & Eddington luminosity and associated error using
$M_{\rm BH}$ estimated from the broad \ion{Mg}{2} line ($L_{\rm Edd}$
= $1.26 \times 10^{38} M_{\rm
  BH}/M_{\rm \odot}$). Units are $L_{\rm \odot}$.\\
lambda\_Edd, lambda\_Edd\_err & Eddington parameter ($\lambda_{\rm Edd}
= L_{\rm AGN}/L_{\rm Edd}$), where $L_{\rm AGN}$ is the bolometric AGN luminosity
estimated from $\lambda L_{\rm 3000}$ and the bolometric correction from
\citet{Runnoe2012}. \\
\enddata
\tablenotetext{1}{We release four catalogs that summarize the properties of the samples analyzed in this work: broad
  \ion{Mg}{2} galaxies (494 sources), SDSS Type 1 quasars (3613) sources,
  SDSS changing-look AGN in the bright state (15 sources), and SDSS
  changing-look AGN in the dim state (11 sources). AGN continuum
  luminosities and emission line properties were measured with
  \textsc{PyQSOFit}.}
\tablenotetext{2}{{\it line} refers to the individual emission lines
  that were fitted: \ion{Mg}{2}\,$\lambda$2780 unresolved doublet, the
  [\ion{Ne}{5}]\,$\lambda$3426
line, the [\ion{O}{2}]\,$\lambda$3728 unresolved doublet, \ion{He}{2}\,$\lambda$4687, H$\beta$
[\ion{O}{3}]\,$\lambda\lambda$5008,4960, H$\alpha$, and [\ion{N}{2}]\,$\lambda\lambda$6549,6585. The suffixes ``br''
refers to a broad line (FWHM $>$ 1200 km s$^{-1}$), ``na'' refers to a narrow
line (FWHM $<$ 1200 km s$^{-1}$), ``c'' refers to the [\ion{O}{3}] line
core, and ``w'' refers to the broadened [\ion{O}{3}] line component
associated with winds. A null value indicates that the line was not
signficantly detected.}
\end{deluxetable*}

When performing a full spectral decomposition that subtracts out both
the host galaxy component and AGN continuum, weak broad Balmer lines
are sometimes significantly detected in the spectra of the broad
\ion{Mg}{2} galaxies. In Table \ref{bl_det}, we quantified the detection of broad Balmer
lines in the sample 
of broad \ion{Mg}{2} galaxies and the comparison samples of SDSS quasars and bright-state and
dim-state changing-look AGN.

It is important to note that broad H$\alpha$ could only be detected in
a subset of sources in each sample due to the long wavelength cutoff
of the SDSS spectrum. SDSS-I and SDSS-II used spectrographs that extended to 9200
\angstrom\ \citep{York2000} while the BOSS spectrographs used in SDSS-III
and SDSS-IV extended to 10400 \angstrom\ \citep{Smee2013}. When
vetting the \textsc{PyQSOFit} results, we retained H$\alpha$ line fit
results when the red end of the spectrum extended beyond the fitted
line center $+ 3\sigma$. With this constraint, we can detect and
reliably fit broad H$\alpha$ up to $z \sim 0.39$ for SDSS-I and
SDSS-II and up to $z \sim 0.57$ for SDSS-III and SDSS-IV. We note in
the parantheses in Table \ref{bl_det} the number of sources in each
sample for which we would be able to detect H$\alpha$ given these
redshift cutoffs. Note that SDSS spectra have low signal-to-noise at
the long wavelength region in which H$\alpha$ can be detected in these
sources. Thus the detection rate of H$\alpha$ may
be driven more by data quality than AGN physics.

\begin{deluxetable}{lll}
\tablecaption{\label{bl_det}Broad Line Detections}
\tablehead{\colhead{Sample} & \colhead{H$\beta$} & \colhead{H$\alpha$\tablenotemark{1}} }
\startdata
Broad \ion{Mg}{2} Galaxies & 57 & 44 (177) \\
SDSS Type 1 Quasars & 2958 & 618 (1268) \\
Bright-State Changing-Look AGN & 15 & 8 (10) \\
Dim-State Changing-Look AGN & 8 &  3 (7) \\
\enddata
\tablenotetext{1}{In parentheses, we note the total number of sources
  per sample in which H$\alpha$ could be detected given the redshift
  of the sources and red wavelength cutoff of the SDSS spectra (9200
  \angstrom\ for SDSS-I and SDSS-II \citep{York2000} and 10400
  \angstrom\ for SDSS-III and SDSS-IV \citep{Smee2013}). Note that at
  the redshifts of these AGN, H$\alpha$ is observed at the longest SDSS
  wavelengths where the spectral S/N is degraded.}
\end{deluxetable}

\subsection{Derived AGN Properties from Spectral Fitting}
From the spectral decomposition, we can use the fitted parameters to
calculate a number of AGN properties, including the AGN continuum
luminosity at 3000 \angstrom\ ($\lambda L_{\rm 3000}$), black hole
mass ($M_{\rm BH}$), and the Eddington ratio ($\lambda_{\rm Edd}$)
which quantifies the accretion rate onto the black hole
relative to the Eddington limit. Using the spectral parameters
reported by \textsc{PyQSOFit}, we calculate these AGN properties for
the broad \ion{Mg}{2} galaxies, Type 1 SDSS quasars, and changing-look
AGN in the bright state and dim state. Here, we hypothesize that the
broadened \ion{Mg}{2} emission in broad \ion{Mg}{2} galaxies is due to
virial gas motion around the black hole such that these spectral
parameters provide information about the fading AGN in the centers of
these galaxies. The average AGN
properties for each sample, and the number of sources anaylzed per
sample, are summarized in Table \ref{agn_prop}.

\subsubsection{Black Hole Mass}
To calculate $M_{\rm BH}$, we use the FWHM of
\ion{Mg}{2} and $\lambda L_{\rm 3000}$ values reported from
\textsc{PyQSOFit} and the single-epoch black hole mass equation from
\citet{Shen2024}\footnote{See the discussion in \citet{Shen2024} about
  their derivation of the single-epoch \ion{Mg}{2}-based black hole
  mass formula which describes why they choise a non-virial exponent
  to the FWHM term. In brief, they find a non-linear relationship
  between the \ion{Mg}{2} line dispersion and FWHM and that their
  chosen exponents for this formula provide more
  consistent black hole mass measurements with those derived via the
  H$\beta$ and \ion{C}{4} lines for SDSS quasars. They also find that
  this formula provides a less biased estimator
  at fixed reverberation-mapped mass than other prescriptions they tried.}:
\begin{align}\label{mbh}
  \rm{log}\left(\frac{M_{\rm BH}}{M_{\rm \odot}}\right) &= -2.05 +
  0.6\times
  \rm{log}\left(\frac{\lambda L_{\rm 3000}}{10^{45}\,\rm{erg}\rm{s}\,^{-1}}\right) +
  \nonumber \\
    &\quad 3\times \rm{log}\left(\frac{\rm{FWHM}}{{\rm km}\,{\rm
  s}^{-1}}\right).
\end{align}
We propagate the uncertainties on $\lambda L_{\rm 3000}$ and
the FWHM of \ion{Mg}{2} to derive the error on $M_{\rm BH}$. We only
retain the measured $M_{\rm BH}$ value when the value exceeds the
error. Compared with the parent samples of broad \ion{Mg}{2} galaxies and SDSS
  quasars, we find similar fractions of sources with
  measured black hole masses within the $0.35 < z < 0.6$ and $0.6 < z
  < 0.75$
  redshift bins, with 51\% (44\%) of broad \ion{Mg}{2} galaxies (SDSS
  quasars) in the low redshift bin.  

For the subset of sources that have broad H$\alpha$ and $\lambda
L_{\rm 5100}$ detections, we independently estimate the
black hole mass using Equation 1 of \citet{greene2010}:
\begin{align}\label{mbhha}
  \rm{log}\left(\frac{M_{\rm BH,H\alpha}}{M_{\rm \odot}}\right) &= 0.807 +
  0.519\times
  \rm{log}\left(\frac{\lambda L_{\rm 5100}}{10^{44}\,\rm{erg}\rm{s}\,^{-1}}\right) +
  \nonumber \\
    &\quad 2.06\times \rm{log}\left(\frac{\rm{FWHM}}{{\rm km}\,{\rm
  s}^{-1}}\right).
\end{align}

\begin{figure}
 \includegraphics[scale=0.55]{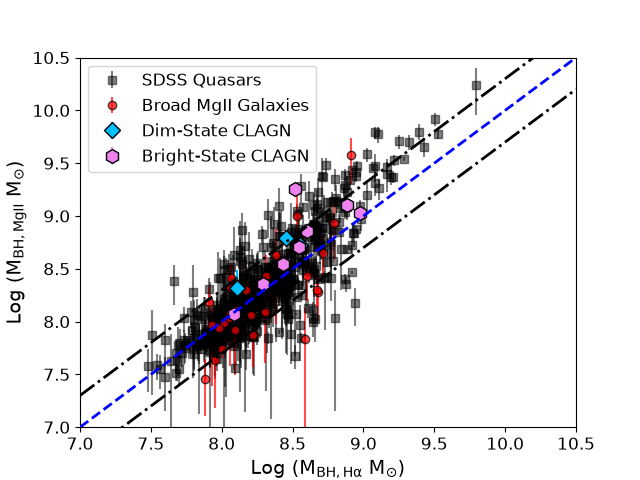}
\caption{\label{comp_mbh} Comparison of black hole masses calculated
  from the broad \ion{Mg}{2} line ($M_{\rm BH,MgII}$) and the broad
  H$\alpha$ line ($M_{\rm BH,H\alpha}$), with a
  dashed blue line indicating the one-to-one relation and a
  dotted-dashed black line denoting the $\pm0.3$ dex error limits
  (i.e., where the independent methods for estimating black hole mass
  agrees to within a factor of 2). The black hole
  masses are consistent for the broad \ion{Mg}{2} galaxy and SDSS
  quasar samples. Though only 5\% of the broad \ion{Mg}{2} galaxy
  sample has estimated black hole masses from the H$\alpha$ line, this
consistency indicates that the broadened \ion{Mg}{2} emission in the
broad \ion{Mg}{2} galaxies is due
to the orbital motion of gas around a supermassive black hole rather
than another physical process.}
\end{figure}

We plot the comparison between $M_{\rm BH}$ (calulated from the
\ion{Mg}{2} line) and $M_{\rm BH,H\alpha}$ in Figure \ref{comp_mbh} for the broad
\ion{Mg}{2} galaxy sample, SDSS quasar sample, and changing-look AGN
samples and report the residuals ($\Delta M_{\rm BH}$  = log($M_{\rm BH}$) - log($M_{\rm
  BH,H\alpha}$)) in Table \ref{agn_prop}.  For the 27 broad \ion{Mg}{2} galaxies where we
had adequate data to calculate $M_{\rm BH}$ and $M_{\rm BH,H\alpha}$, we find consistent black hole
masses with the two methods within 0.04 dex.

A comparison of the \ion{Mg}{2} single-epoch mass
estimator from \citet{Shen2024} with the H$\alpha$ single-epoch mass
formula from \citet{greene2010} was performed by \citet{lamassa2024}
using quasars from the Stripe 82X survey
\citep{lamassa2013a,lamassa2013b,lamassa2016b} that had SDSS spectra. 
Only 12 quasars lie within the appropriate redshift
range for both \ion{Mg}{2} and H$\alpha$ to be covered in the spectrum
and had these lines significantly detected, but the results in
\citet{lamassa2024} are similar to those
reported here: the masses between these two methods agree to within an
average value of 0.07 dex and, as seen in Figure \ref{comp_mbh}, the \ion{Mg}{2} derived black hole
  masses are systematically higher than the H$\alpha$ derived black
  hole masses at high mass \citep[i.e., $> 3\times10^{8}$ M$_{\sun}$,
  see their Figure 5,][]{lamassa2024}.

The agreement between the \ion{Mg}{2}-derived and H$\alpha$ derived
black hole masses in both the broad \ion{Mg}{2} galaxy sample and comparison quasar and changing-look AGN samples is compatible with the hypthosesis that the broadened
\ion{Mg}{2} emission in the broad \ion{Mg}{2} galaxies is due to the virial motion of
gas orbiting a
supermassive black hole hole rather than another physical process,
like starburst powered outflows.

\subsubsection{Accretion Rate}
We use $M_{\rm BH}$ to calculate the Eddington luminosity ($L_{\rm
  Edd}$) which is the theoretical maximum luminosity of an object
when the gravitational force inward is balanced by radiation pressure
outward. $L_{\rm Edd}$ is defined as $1.26 \times 10^{38} M_{\rm
  BH}/M_{\rm \odot}$. We propagate the
uncertainties on the black hole mass to
calculate the error on $L_{\rm Edd}$. We estimate the bolometric AGN luminosity
($L_{\rm AGN}$) by
applying the bolometric correction of $5.18\pm0.19$ to $\lambda L_{\rm
  3000 \angstrom}$ from \citet{Runnoe2012}. 

Finally, we calculate the Eddington
parameter: $\lambda_{\rm Edd}= L_{\rm AGN}/L_{\rm Edd}$.  We propagate the uncertainties on the bolometric correction,
$\lambda L_{\rm 3000}$, and $L_{\rm Edd}$ to calculate the
error on $\lambda_{\rm Edd}$. Again, we retain $\lambda_{\rm Edd}$
measurements for the sources where the value exceeds the error. While we have measurements of $\lambda L_{\rm 3000}$ for all
sources in the broad \ion{Mg}{2} galaxy, SDSS quasar, and changing-look AGN
samples (by design), our requirement that we have well constrained
measurements for $M_{\rm BH}$ and $\lambda_{\rm Edd}$ restricts our
comparison of these parameters to 66\%, 93\%, 100\%, and 91\% of these
samples, respectively (see Table \ref{agn_prop}); the fractions
  of broad \ion{Mg}{2} galaxies and SDSS quasars with $\lambda_{\rm
    Edd}$ measuresments in the low and high
  redshift bins are identical to the fractions with $M_{\rm BH}$
  measurements in these bins.

\begin{deluxetable}{llllll}
\tablecaption{\label{agn_prop}Average AGN Properties\tablenotemark{1}}
\tablehead{\colhead{Sample} & \colhead{log($\lambda L_{\rm
      3000}$, erg s$^{-1}$)}  & \colhead{log($L_{\rm MgII}$, erg s$^{-1}$)} 
    & \colhead{log($M_{\rm BH}$, M$_{\rm \odot}$)\tablenotemark{2}}  &
    \colhead{log($\lambda_{\rm Edd}$)} & \colhead{$\Delta M_{\rm
        BH}$\tablenotemark{3}} \\
  & \colhead{(dex)} & \colhead{(dex)} & \colhead{(dex)} &
  \colhead{(dex)} & \colhead{(dex)} }
\startdata
Broad \ion{Mg}{2} Galaxies & $43.7\pm0.3$ & $42.3\pm0.2$ &  $8.35\pm0.55$ &
$-2.1\pm0.5$ & $-0.04 \pm 0.30$\\
Sample Size & 494 & 494 & 325 & 322 & 27\\
\hline
SDSS Quasars & $44.4\pm0.5$ & $42.7\pm0.4$ & $8.52\pm0.50$ & $-1.5\pm0.5$ & $0.04\pm0.25$\\
Sample Size & 3613 & 3613& 3375 & 3374 & 565 \\
\hline
Bright-State Changing-Look AGN & $44.6\pm0.4$ & $43.0\pm0.4$ & $8.81\pm0.43$ &
$-1.6\pm0.4$ & $0.20\pm0.22$\\
Sample Size & 15 & 15 & 15 & 15 & 8 \\
\hline
Dim-State Changing-Look AGN & $44.2\pm0.5$ & $42.9\pm0.4$ & $8.83\pm0.43$ &
$-2.0\pm0.6$ & $0.24\pm0.07$ \\
Sample Size & 11 & 11 & 10 & 10 & 3 \\
\enddata
\tablenotetext{1}{The reported error bars refer to the standard deviation.}
\tablenotetext{2}{$M_{\rm BH}$ is calculated using the FWHM of the
  \ion{Mg}{2} line and $\lambda L_{\rm3000}$ continuum luminosity (see
  Equation \ref{mbh}).}
\tablenotetext{3}{$\Delta M_{\rm BH}$  = log($M_{\rm BH}$) - log($M_{\rm
  BH,H\alpha}$), where $M_{\rm BH}$ is defined above and $M_{\rm
  BH,H\alpha}$ is calculated using the FWHM of H$\alpha$ and $\lambda
L_{\rm 5100}$ (see Equation \ref{mbhha}).}
\end{deluxetable}

\section{Discussion}

\subsection{Changing-Look AGN Comparison}
\begin{figure*}
\includegraphics[scale=0.55]{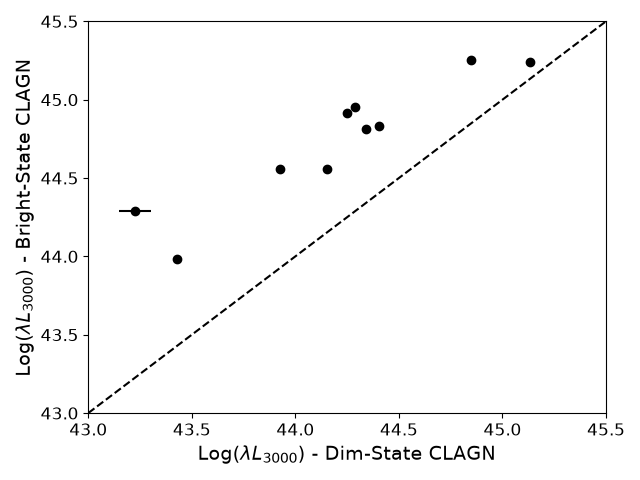}~
 \includegraphics[scale=0.55]{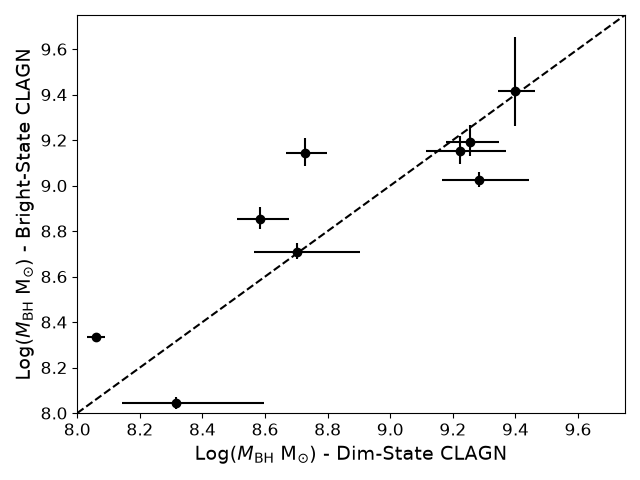}
\caption{\label{comp_clagn} Comparison of AGN continuum luminosity
  (left) and black hole mass (right) for changing-look AGN between
  their bright state and dim state. The dashed line in both plots mark
where these quantities are equal. As expected, 
changing-look AGN have higher $\lambda L_{\rm 3000}$ values in their
bright state than in their dim state. Their black hole mass
measurements from both spectral epochs are consistent and agree on
average to within a factor of two.}
\end{figure*}

We compare the continuum luminosity  ($\lambda L_{\rm 3000}$) and black
hole masses for changing-look AGN where we have these measurements for
the same sources in both the bright and dim states. As shown in Figure
\ref{comp_clagn} (left), the AGN continuum luminosity is
systematically higher in the bright state than the dim state, which is
expected. We compare the black hole masses in the right hand panel of
Figure \ref{comp_clagn} and find that the mass measurements are
comparable in the dim state and bright state, with an average offset
of 0.04 dex and standard deviation of 0.23 dex. Though the AGN continuum
luminosity is weaker in the dim state, the increase in the \ion{Mg}{2}
FWHM compensates so that the black hole mass measurements from both
spectral epochs are roughly equal.

\subsection{Broad \ion{Mg}{2} Galaxies on the AGN Continuum}
\begin{figure*}
\includegraphics[scale=0.55]{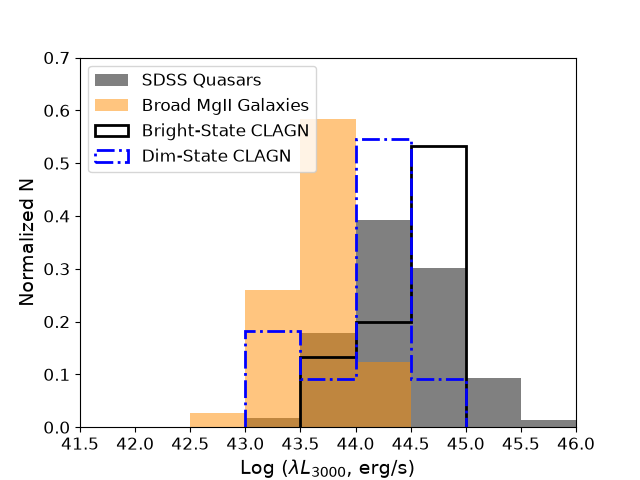}~
 \includegraphics[scale=0.55]{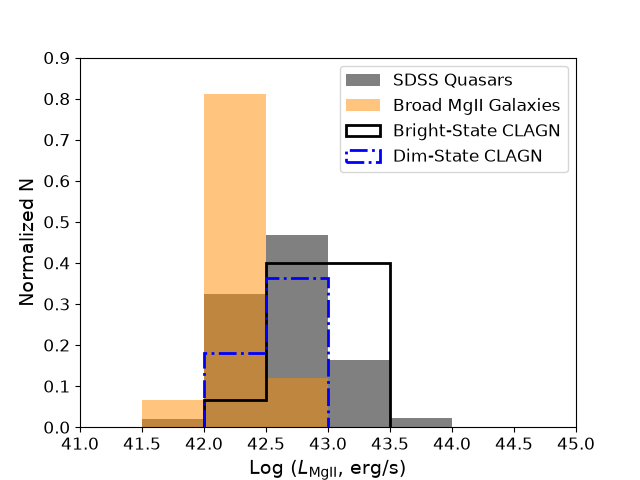}
\includegraphics[scale=0.55]{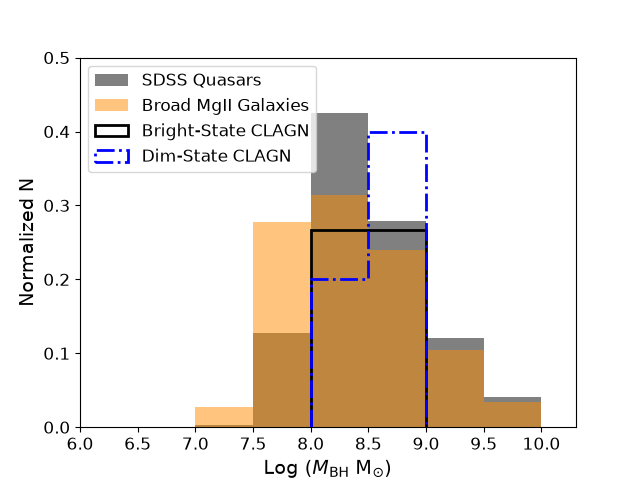}~
\includegraphics[scale=0.55]{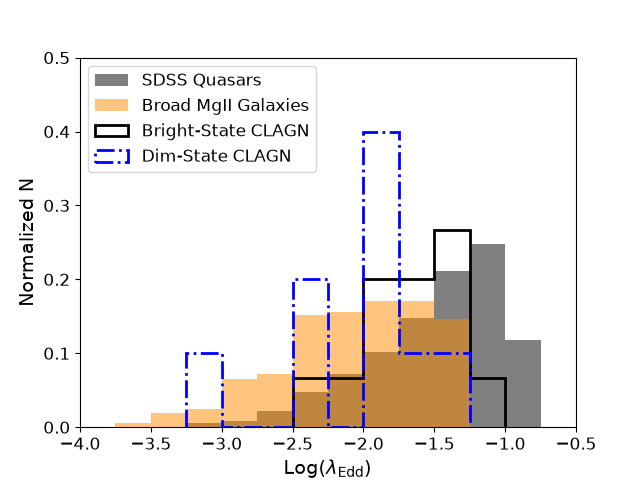}
\caption{\label{agn_prop_hist} Distribution of (top left) AGN
  continuum luminosity ($\lambda L_{\rm 3000}$), (top right) broad
  \ion{Mg}{2} luminosity ($L_{\rm MgII}$),  (bottom left)
$M_{\rm BH}$,  and (bottom right) Eddington parameter ($\lambda_{\rm
  Edd}$) for the broad \ion{Mg}{2} galaxies (solid orange), SDSS quasars (solid
grey), bright-state changing-look AGN (black solid line), and
dim-state changing-look AGN (blue dot-dashed line). While the broad
\ion{Mg}{2} galaxies have black hole masses that are on average similar to typical
quasars and changing-look AGN, they have lower \ion{Mg}{2} and AGN continuum
luminosities. Both dim-state changing-look AGN and broad \ion{Mg}{2}
galaxies tend to have lower $\lambda_{\rm Edd}$ values compared with
SDSS quasars and bright-state changing-look AGN which indicates that a
suppressed accretion rate is responsible for the weak broad Balmer
lines in these galaxies. Normalized
distributions are shown to ease visual comparison due to disparate
sample sizes.}
\end{figure*}

Figure \ref{agn_prop_hist} shows the distribution of $\lambda L_{\rm 3000}$,
$M_{\rm BH}$, and $\lambda_{\rm Edd}$. In Table \ref{agn_prop_stats}, we report the
results of the k-sample Anderson-Darling test \citep{adtest} with
permutation, comparing these properties between the
broad \ion{Mg}{2} and the AGN samples to quantify whether they are drawn
from statistically different parent populations. Due to the disparate
sample sizes, spanning 1-2 orders of magnitude, we use the
k-sample Anderson-Darling test rather than the often
used Kolmogorov–Smirnov test since
Anderson-Darling is 
less sensitive to disparate samples sizes and is more sensitive to
differences in the tails of the distibutions. We use permutation,
drawing $5\times10^3$ samples from a random distribution, to
calculate the $p$-value which offers more accurate $p$-values for
small sample sizes than using a $p$-value that is interpolated from
tabulated values (i.e., the default, non-permutation method). We simply use the $p$-value to report whether the
Anderson-Darling test indicates that the distribution of AGN
properties are significantly different between pair-wise comparisons
of samples considered in this work, where we use $p < 0.01$ as a
metric to conclude that the distributions are drawn from different
parent populations. The statistical
test results reported in Table \ref{agn_prop_stats} agree with a qualitative comparison of
average AGN properties reported in Table \ref{agn_prop}, which we
discuss in further detail below.

\begin{deluxetable*}{lllll}
\tablecaption{\label{agn_prop_stats}Results of k-sample
  Anderson-Darling Test\tablenotemark{1}}
\tablehead{\colhead{Pair-Wise Sample Comparison} &
  \colhead{$\lambda L_{\rm
      3000}$ $p$-value} & \colhead{$L_{\rm
      MgII}$ $p$-value} 
    & \colhead{$M_{\rm BH}$ $p$-value}  &
    \colhead{$\lambda_{\rm Edd}$ $p$-value}}
\startdata
Broad \ion{Mg}{2} Galaxies \& & $2\times10^{-4}$ & $2\times10^{-4}$ &
$2\times10^{-4}$ & $2\times10^{-4}$ \\
SDSS Quasars \\
Broad \ion{Mg}{2} Galaxies \& & $2\times10^{-4}$ &
$2\times10^{-4}$ & $2\times10^{-3}$ & $2\times10^{-4}$ \\
Bright-State Changing-Look AGN \\
Broad \ion{Mg}{2} Galaxies \& & $2\times10^{-4}$ &
$2\times10^{-4}$ & $8\times10^{-3}$ & 0.68 \\
Dim-State Changing-Look AGN \\
\hline
SDSS Quasars \& & 0.07 & $2\times10^{-4}$ & 0.01 & 0.54 \\
Bright-State Changing-Look AGN \\
SDSS Quasars \& & 0.22 & 0.04 & 0.04 &
$2\times10^{-3}$ \\
Dim-State Changing-Look AGN \\
\hline
Bright-State Changing-Look AGN \& & 0.02 & 0.72 & 
0.78 & 0.13 \\
Dim-State Changing-Look AGN \\
\enddata
\tablenotetext{1}{We used the \textsc{scipy.stats.anderson\_ksamp}
  routine with permutation to calculate $p$-values, which tests
  whether two samples are drawn from the same parent
  distribution. A $p$-value $< 0.01$
indicates that we can reject the null hypothesis that the two
distributions are drawn from the same parent population.}
\end{deluxetable*}

The Anderson-Darling test indicates that the distribution of black
hole masses between SDSS Type 1 quasars and broad \ion{Mg}{2} galaxies
are drawn from different parent populations, though the average black
hole masses between the two samples are not significantly different (average log($M_{\rm BH}/M_{\rm
  \odot}$) = $8.35 \pm 0.55$ dex compared with
$8.52\pm0.50$ dex for SDSS quasars).

We find larger disagreements between the
average AGN luminosities and Eddington parameters between broad
\ion{Mg}{2} galaxies and typical SDSS quasars (average factors of 0.7
dex and 0.6 dex, respectively), which would be expected if the broad
\ion{Mg}{2} galaxies are hosting fading AGN. In fact, the broad
\ion{Mg}{2} galaxies have the lowest average AGN luminosity of all
samples analyzed in this work, with underlying luminosity
distributions that are significantly different from all AGN samples we
analyzed. This result could be understood in the context of the
selection of broad
\ion{Mg}{2} galaxies. These
sources are spectroscopically identified as ``Galaxies'' and careful
spectroscopic decomposition is required to reveal their putative AGN
nature, so their AGN luminosity could be expected to be weaker than
spectroscopically identified AGN.

How can we
test whether apparent weak AGN emission in these
galaxy dominated spectra is actually due to an AGN and not a relic of
overfitting spectra to extract a weak AGN component? Here we use the luminosity of the broad \ion{Mg}{2} emission line as a
clue. We posit that the broad \ion{Mg}{2} line in these galaxies is
powered by the AGN continuum as it is in bona fide AGN. However,
the \ion{Mg}{2} line has a lower responsivity than the Balmer lines
\citep{KoristaGoad2004}, so we may expect the broad \ion{Mg}{2} line
luminosity ($L_{\rm MgII}$) to be a lagging indicator of the instantaneous AGN
luminosity though still a proxy of the AGN continuum luminosity.

With this caveat in mind, we compare the distributions of $L_{\rm MgII}$ among the broad \ion{Mg}{2} galaxies, SDSS Type 1
quasars, and changing-look AGN we consider in this work (Figure
\ref{agn_prop_hist}, top right)
and report the average  $L_{\rm MgII}$ values for each sample
in Table \ref{agn_prop}.  We find that, similar to the comparison of
the AGN continuum luminosity distributions, the broad \ion{Mg}{2} galaxies
have the lowest $L_{\rm MgII}$ values amongst the samples we considered and that the
parent distribution of $L_{\rm MgII}$ values for the broad \ion{Mg}{2}
galaxies is significantly different than those of the bona fide
AGN. We thus conclude that the weaker \ion{Mg}{2} luminosity is due to
the AGN being weaker and that the spectral decomposition performed by
\textsc{PyQSOFit} extracts
this otherwise hidden AGN continuum luminosity. We caution that
  due to the stringent cuts to extract significant broad \ion{Mg}{2}
  line emission from the spectra, these $L_{\rm MgII}$ distributions
  do not represent the intrinsic $L_{\rm MgII}$ distribution of these
  populations and what is shown here is by design biased to higher
  luminosity values. However, as these cuts were uniformly applied to
  all samples, the results of the intercomparison across these
  populations does not depend on the intrinsic distribution.

\subsection{Eddington Ratio as a Driver of AGN Evolution}

Similar to
results from past studies of extremely variable quasars \citep{Rumbaugh2018} and
changing-look AGN \citep{MacLeod2019,Graham2020,Green2022,Zeltyn2024}, $\lambda_{\rm Edd}$ is systematically lower
for dim-state changing-look AGN than Type 1 quasars, and lower for dim-state
changing-look AGN than bright-state changing-look AGN
\citep{Wang2024,Guo2025,Dong2025}, though the
distributions of $\lambda_{\rm Edd}$ are not significantly
different between the small samples of bright- and dim-state changing
look AGN considered here. We find that the distribution of
$\lambda_{\rm Edd}$ and the average value of $\lambda_{\rm Edd}$ are similar between changing-look
AGN in the dim state and broad \ion{Mg}{2} galaxies. Taking these
results together, we see that broad \ion{Mg}{2} galaxies lie on the
tail end of the continuum of AGN activity with an Eddington ratio
below that of typical Type 1 quasars and similar to that of dim-state
changing look AGN (see Tables \ref{agn_prop}-\ref{agn_prop_stats}).

It has been proposed that a decrease in accretion rate drives an
evolution in AGN spectral type from Type 1.0 $\rightarrow$ 1.n $\rightarrow$
2.0 \citep{Elitzur2014}. Due to H$\beta$ having a greater
responsivity to the ionizing continuum luminosity than H$\alpha$,
broad H$\beta$ gradually fades followed by broad H$\alpha$ as the AGN
continuum dims \citep{KoristaGoad2004}. Many studies of
changing-look AGN invoke this framework to describe the
(dis)appearance of broad Balmer lines and weakening/strengthening of
the AGN continuum over time
\citep[e.g.,][]{lamassa2015,runnoe2016,MacLeod2016,MacLeod2019,Green2022}.

Where does broad \ion{Mg}{2} lie on this sequence? \ion{Mg}{2} is a
collisionally excited line and
has a lower responsivity than the photoionized Balmer lines \citep{Guo2020a}. Observational
studies find that \ion{Mg}{2} displays less variability than broad
H$\beta$ in both typical quasars \citep{Kokubo2014,sun2015} and changing-look
AGN \citep{Zeltyn2024,zeltyn2026} and that the broad \ion{Mg}{2} line has
longer reverberation time lags than the broad H$\beta$ line
\citep{homayouni2020}, consistent with theoretical
predictions about its lower responsivity. In particular, the
  recent study by \citet{zeltyn2026} found that broad \ion{Mg}{2}
  emission remains prominent in the majority of dimming changing-look
  AGN where broad H$\beta$ emission disappeared (where this
  measurement was feasible). We could thus expect that 
  broad \ion{Mg}{2} galaxies represent a phase of weakened AGN
  activity in the changing-look AGN sequence where broad \ion{Mg}{2} emission endures after the broad Balmer lines fade.

Previous work has drawn an analogy
between changing-look AGN and state transitions in stellar mass X-ray
binaries where the inner accretion disk evaporates and is replaced
by an advection dominated accretion flow which occurs at an Eddington
ratio of a few percent \citep[e.g.,][]{maccarone2003,done2007}. Observational results
lend support to this model with X-ray spectral states of changing-look AGN
correlating with Eddington ratio changes in a manner similar to X-ray
binaries \citep{Noda2018,Ruan2019,ricci2023,jana2025,jana2026}, UV-optical
spectral energy distributions of dim-state changing look AGN being
well described by an advection-dominated accretion flow and truncated disc model \citep{duffy2025}, and dim-state changing look AGN
being predomoninantly at Eddington ratios below a threshold of
$\sim$0.01 \citep{Wang2024}. In our analysis, we find that the average
Eddington ratio is at this transition threshold for both broad \ion{Mg}{2}
galaxies (log($\lambda_{\rm Edd}$) = $-2.1 \pm 0.5$) and dim-state
changing-look AGN (log($\lambda_{\rm Edd}$) = $-2.0 \pm 0.6$),
  though the range in average $\lambda_{\rm Edd}$ values is large and
  there is no evidence of a sharp transition at $\lambda_{\rm Edd}
  \sim 0.01$. However,  given the similarities in AGN luminosity and
  Eddington ratio distributions between broad \ion{Mg}{2} galaxies and
  dim-state changing look AGN, it seems plausible that the same
  physical driver, i.e., low accretion rates, power the broad emission lines in both populations.

Broad \ion{Mg}{2} galaxies can represent  a penultimate phase in the dimming sequence of changing-look AGN.
When the Eddington ratio drops below a critical
threshold that is perhaps associated with an accretion state transition, the AGN continuum and broad Balmer lines weaken,
manifesting as an evolution in AGN spectral type observed in
changing-look AGN. During the transition phase, an AGN can become a dim-state
changing-look AGN and then a broad \ion{Mg}{2} galaxy, where the broad
\ion{Mg}{2} line remains prominent due to its
the lower responsivitiy compared with the Balmer lines. In some
changing-look AGN, the broad \ion{Mg}{2} line weakens \citep[see,
e.g.,][]{guo2019,guo2024} which could perhaps be driven by a further
decrease in the Eddington ratio. Indeed, by design we only
  included changing-look AGN in this analysis where broad \ion{Mg}{2}
  emission was significantly detected to test the hypothesis that
  broad \ion{Mg}{2} AGN are dim-state changing-look AGN at the next
  stage of evolution. Further down the sequence are changing-look AGN
  where the broad \ion{Mg}{2} line is absent which is a population not
  considered in this work. In such sources, whose spectra would appear
  as typical Type 2 AGN, care is needed to identify reasonable proxies of
  the intrinsic AGN continuum and Eddington ratio. Ideally the same
  proxies should be used for Type 1 AGN and
broad \ion{Mg}{2} galaxies to enable a fair intercomparison of AGN
luminosity and Eddington ratio distributions to assess where these populations lie
on the proposed AGN evolutionary sequence.

\subsection{Baldwin Effect}
The Baldwin Effect is an observed anticorrelation between the
equivalent width (EW) of broad lines and the AGN continuum luminosity
\citep{Baldwin1977}. \citet{Dietrich2002}  demonstrated that the strength of the anticorrelation (i.e.,
 slope) between
 the EW and AGN continuum luminosity depends on the emission line,
 where \ion{Mg}{2} had a
 strong anti-correlation, but there was no significant Baldwin effect for H$\beta$ \citep[i.e., the fitted
 slope was consistent with 0, though see][]{netzer2007}. If broad \ion{Mg}{2} galaxies are fading AGN where broad
\ion{Mg}{2} emission is a signpost of low levels of black hole accretion, we may therefore
expect to observe a Baldwin effect between \ion{Mg}{2} and the AGN
continuum luminosity at $\lambda L_{\rm 3000}$. For the
sources where broad Balmer lines are detected (i.e., H$\beta$ and H$\alpha$), we would also expect
that any relation between the broad lines and the AGN continuum
powering these lines
(e.g., $\lambda L_{\rm 5100}$) would be weaker than that observed
between \ion{Mg}{2} and $\lambda L_{\rm 3000}$.

\begin{figure*}
  \begin{center}
 \includegraphics[scale=0.55]{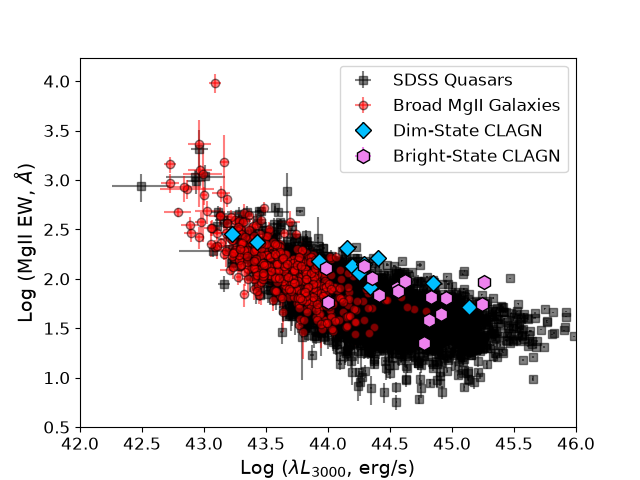}~
\includegraphics[scale=0.55]{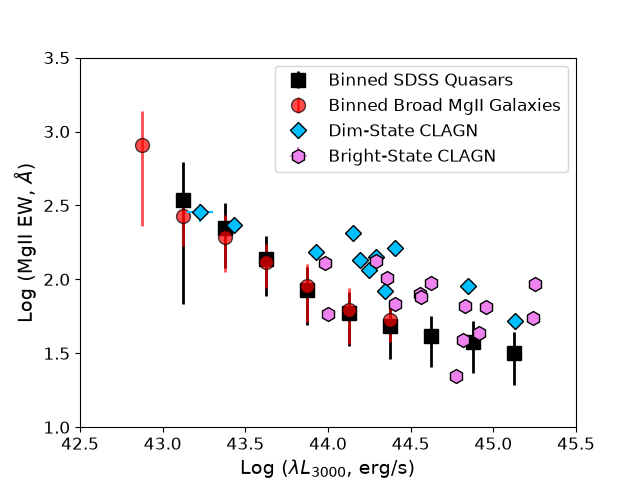}
\caption{\label{mgii_baldwin} Broad \ion{Mg}{2} EW as a function of
  AGN continuum luminosity at 3000 \angstrom\ for Type 1 quasars from
  SDSS (black squares), broad \ion{Mg}{2} galaxies (red circles),
  dim-state changing-look AGN (blue diamonds), and bright-state
  changing-look AGN (purple hexagons). As reported in Table
  \ref{baldwin_effect}, we find a clear Baldwin effect with strong correlations (Pearsons $|r| >
  0.5$) for the SDSS
  quasars, broad \ion{Mg}{2} galaxies, and dim-state changing-look
  AGN. We show binned versions of the \ion{Mg}{2} EW vs $\lambda L_{\rm
    3000 \angstrom}$ relationship for the SDSS
  quasars and broad \ion{Mg}{2} galaxies in the right panel for
  visualization.  }
\end{center}
\end{figure*}

\begin{figure*}
  \begin{center}
 \includegraphics[scale=0.55]{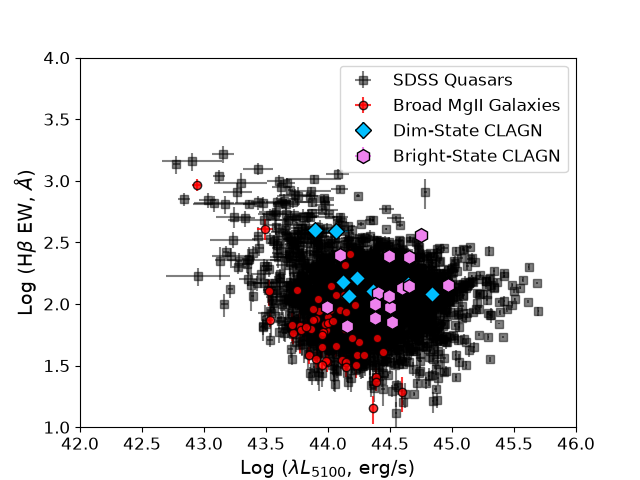}~
\includegraphics[scale=0.55]{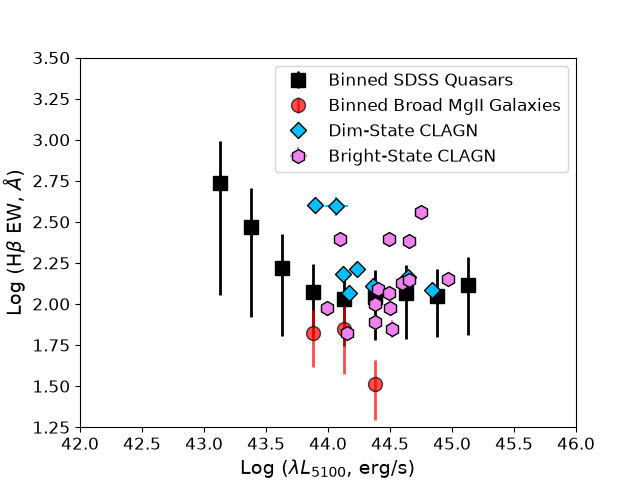}
\caption{\label{hb_baldwin} Same as Figure \ref{mgii_baldwin} but with
the broad H$\beta$ line as a function of the AGN continuum luminosity
at 5100 \angstrom. There are no strong correlations for the broad
\ion{Mg}{2} galaxy, SDSS quasars and bright-state changing-look AGN
samples. Though the $r$ value for the dim-state changing-look AGN
sample suggests a strong correlation, the uncertainty on the fitted
slope gives a shallower relationship than between the \ion{Mg}{2} EW
and $\lambda L_{\rm 3000}$.}
\end{center}
\end{figure*}

\begin{figure*}
  \begin{center}
 \includegraphics[scale=0.6]{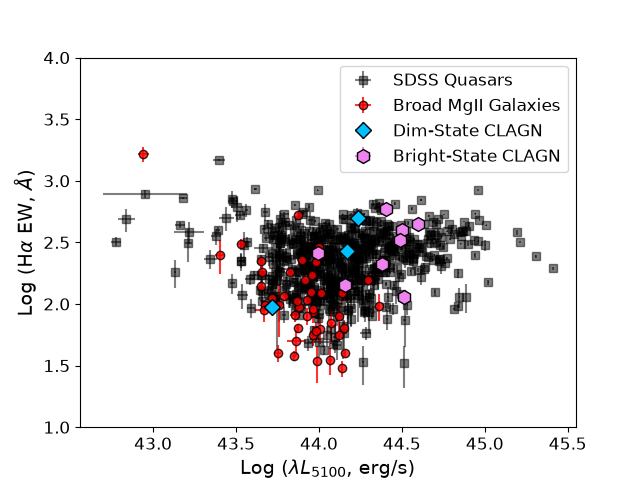}~
\includegraphics[scale=0.6]{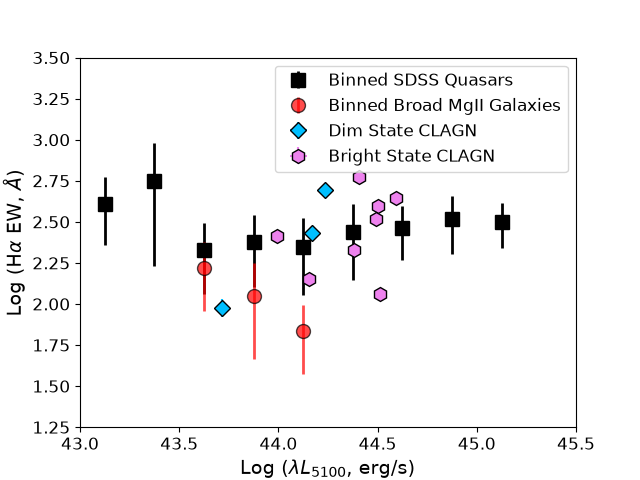}
\caption{\label{ha_baldwin} Same as Figure \ref{hb_baldwin} but for
  H$\alpha$. The correlation coefficient suggests a strong correlation
for the broad \ion{Mg}{2} galaxy sample, but the correlation and
linear relationship (i.e., fitted slope) is much weaker than observed
with the relationship between broad \ion{Mg}{2} EW and $\lambda L_{\rm
  3000}$. }
\end{center}
\end{figure*}

We assess the strength of the correlation between the broad line EW
and AGN continuum luminosity using the Pearson correlation coefficient
($r$), where we remove outliers when the sample size is $\geq$50 by
performing an interative linear least square fitting with a 3$\sigma$
clipping.\footnote{For this analysis, we used \textsc{astropy}'s
  \textsc{LinearLSQFitter} fitter and
  \textsc{FittingWithOutlierRemoval} routine. We used
  \textsc{scipy.stats.pearsonr} to calculate the correlation coefficient.}
  In Table \ref{baldwin_effect}, we report both $r$ and the
fitted slope ($m$) between the broad line EW and AGN continuum
luminosity, where the number of sources in each sample are
reported. For this excercise, we require the error on $\lambda L_{\rm
  5100}$ to be below 0.3 dex when comparing this luminosity
with the broad H$\beta$ and H$\alpha$ EW values. In Figures \ref{mgii_baldwin}-\ref{ha_baldwin}, we plot the \ion{Mg}{2}, H$\beta$, and
H$\alpha$ EW (respectively) as a function of the AGN continuum
luminosity (left panels) for the broad \ion{Mg}{2}, SDSS quasar, and
changing-look AGN samples. In the right panels of Figures \ref{mgii_baldwin}-\ref{ha_baldwin}, we
show for visualization purposes binned versions of these plots, where:
the luminosity bin size is 0.25 dex, we required at least 5 objects per bin,
and we calculated the 3$\sigma$-clipped mean of the EW and standard deviation
 for each binned datapoint;\footnote{We used
   \textsc{astropy.stats.sigma\_clipped\_stats} to calculate the
   sigma-clipped mean and standard deviation.} we do not bin the changing-look AGN values
 since there are $\leq$15 sources in these samples.

 Similar to the Type 1 AGN analyzed by \citet{Dietrich2002} and the SDSS quasars analyzed here,
 we find that broad \ion{Mg}{2} galaxies have a strong Baldwin effect
 in \ion{Mg}{2} (i.e., $r = -0.76$, $m = -0.55 \pm 0.02$) but not in the
 Balmer lines. The fitted slope between H$\beta$ and the AGN
 continuum luminosity is nearly flat (when considering the
 uncertainty) and trends in the opposite direction expected. While
 $|r| > 0.5$ for the H$\alpha$ EW vs $\lambda_{\rm 5100}$
 relationship, the slope is much more shallow than the one for the
 \ion{Mg}{2} vs $\lambda_{\rm 3000}$ relationship. The
 changing-look AGN also show a stronger \ion{Mg}{2} Baldwin Effect
 than the Balmer lines.

 These results agree qualitatively with the framework described by \citet{Guo2020a} who performed locally
 optimally emitting cloud (LOC) photionization modeling to describe
 the behavior of \ion{Mg}{2} and the broad Balmer lines when the AGN
 continuum luminosity varies. While the models
 published in the \citet{Guo2020a} study probe AGN luminosities beyond those of
 our broad \ion{Mg}{2} galaxy sample (i.e., \citet{Guo2020a} model AGN
 where $\lambda L_{\rm 3000} >
 10^{44}$ erg s$^{-1}$), they extrapolate
 their LOC models to predict broad the \ion{Mg}{2} EW as a function of
 $\lambda L_{\rm 3000}$ from $10^{43}$ erg s$^{-1} < \lambda
  L_{\rm 3000} < 10^{45}$ erg s$^{-1}$ and they report a good
  match to the measured properties of the broad \ion{Mg}{2} galaxies
  reported in \citet{roig2014}. Though we are unable to directly test
  these model predictions on our sample of broad \ion{Mg}{2} galaxies,
  the qualitative agreement between our results and those discussed in
  \citet{Guo2020a} are consistent with the hypothesis that broad
  \ion{Mg}{2} galaxies are part of the changing-look AGN sequence.

\begin{deluxetable*}{lllllll}
\tablecaption{\label{baldwin_effect}Baldwin Effect: Correlations
  between broad line EW and AGN Continuum Luminosity\tablenotemark{1}}
\tablehead{
  \colhead{Sample} & 
  \multicolumn{2}{c}{\ion{Mg}{2} EW vs $\lambda L_{\rm 3000
      \angstrom}$} & 
  \multicolumn{2}{c}{H$\beta$ EW vs $\lambda L_{\rm 5100}$} &
  \multicolumn{2}{c}{H$\alpha$ EW vs $\lambda L_{\rm 5100}$} \\
  & \colhead{$r$} & \colhead{$m$} &
   \colhead{$r$} & \colhead{$m$} &
    \colhead{$r$} & \colhead{$m$}
}
\startdata
Broad \ion{Mg}{2} Galaxies & $-0.76$ & $-0.55 \pm 0.02$ & $-0.35$ & $0.18 \pm 0.14$ & $-0.57$ & $-0.41 \pm 0.26$ \\
Sample Size & 488 & & 55 & & 45 & \\
\tableline
SDSS Quasars & $-0.62$ & $-0.22 \pm 0.01$ & $-0.09$ & $-0.01 \pm 0.01$ &
0.09 & $0.11 \pm 0.02$ \\
Sample Size & 3567 & & 2800 & & 608 & \\
\tableline
Bright-State Changing-Look AGN & $-0.40$ & $-0.17 \pm 0.12$ & 0.35 &
$0.43 \pm 0.30$ & 0.30 & $0.56 \pm 0.42$ \\
Sample Size & 15 & & 15 & & 8 & \\
\tableline
Dim-State Changing-Look AGN\tablenotemark{2} & $-0.89$ & $-0.43 \pm 0.06$ & $-0.65$ &
$-0.38 \pm 0.27$ & \nodata & \nodata \\
Sample Size & 11 & & 8 & & 3 & \\
\enddata
\tablenotetext{1}{$r$ refers to the Pearson correlation coefficient
  and $m$ is the fitted slope where the  AGN luminosity is the independent variable and broad line EW is
  the dependent variable.}
\tablenotetext{2}{With only 3 dim-state Changing-Look AGN that have
  broad H$\alpha$ detections, we do not calculate the correlation
  coefficient or slope between broad H$\alpha$ EW and $\lambda L_{\rm
    5100 \angstrom}$ for these objects.}
\end{deluxetable*}

\subsection{Ionization Hardness from the narrow [\ion{O}{3}]/H$\beta$ ratio}
Type 2 AGN that lack broad emission lines can be identified by the
ratios of narrow emission lines, where the BPT diagram is the most
often used diagnostic \citep{bpt,vo1987,Kewley2001,Kauffmann2003}. This diagnostic uses a combination
of the narrow [\ion{O}{3}]/H$\beta$ and [\ion{N}{2}]/H$\alpha$ flux ratios
to identify galaxies harboring accreting black holes.\footnote{These
  ratios use the [\ion{O}{3}] transition at 5007 \angstrom\ and
  [\ion{N}{2}] transition at 6584 \angstrom.}  [\ion{O}{3}]/H$\beta$ traces the
ionization field hardness while [\ion{N}{2}]/H$\alpha$ is a proxy of the
metallicity. Due the redshift of our sample, [\ion{N}{2}]/H$\alpha$ is
mostly unavailable to us, and we do not have stellar masses of the AGN
host galaxies to use the alternate Mass Excitation diagnostic
\citep[MEx,][]{juneau2011}, but we investigate whether the  [\ion{O}{3}]/H$\beta$
ratio supports the idea that broad \ion{Mg}{2} galaxies harbor AGN.

As
the narrow line region, which is photoionized by the AGN, extends
hundreds to thousands of parsecs beyond the obscuring torus, the ratio
of  [\ion{O}{3}]/H$\beta$ can serve as a relic of more active accretion
past while the broad line region responds to AGN continuum dimming
on shorter timescales. A high [\ion{O}{3}]/H$\beta$ ratio is not a
definitive diagnostic of AGN ionization, however, as low metallicity galaxies can have
higher [\ion{O}{3}]/H$\beta$ ratios than galaxies at
solar metallicty. Conversely, dust and host galaxy dilution can
attenuate optical emission lines, hiding signatures of the AGN
ionizing continuum in an optical spectrum.

\begin{figure}
  \begin{center}
 \includegraphics[scale=0.55]{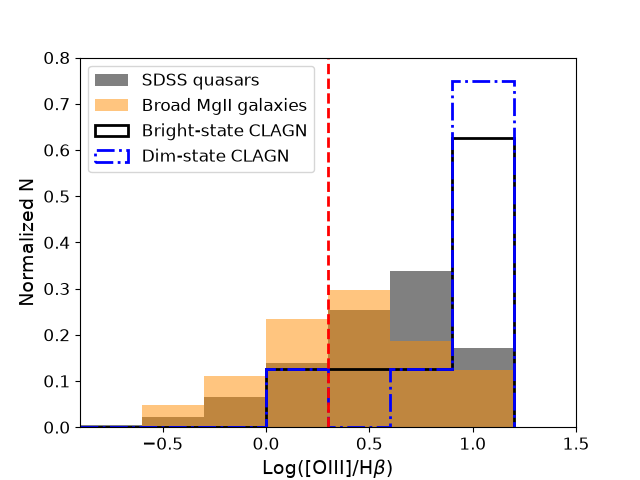}
\caption{\label{o3_hb_hist} Distribution of [\ion{O}{3}]/H$\beta$, a
  proxy of the ionization field hardness, for
  the broad \ion{Mg}{2}, SDSS quasar, and changing-look AGN
  samples. The red dashed line denotes the log([\ion{O}{3}]/H$\beta$) $>$
0.3 dex threshold above which gas is primarily ionized by AGN
accretion for galaxies that are not low metallicty
\citep{zakamska2003,reyes2008}. A majority of the sources in each
sample are above this threshold (see Table \ref{o3_hb_tab}). }
\end{center}
\end{figure}

We detect [\ion{O}{3}] and narrow H$\beta$ for 51\%, 47\%, 53\%,
and 73\% of the broad \ion{Mg}{2}, SDSS quasar, bright-state
changing-look AGN, and dim-state changing-look AGN samples,
respectively (see Table \ref{o3_hb_tab}). We plot the distribution of this ratio for these samples
in Figure \ref{o3_hb_hist}. As reported in Table \ref{o3_hb_tab}, we find that the changing-look AGN samples have the
highest [\ion{O}{3}]/H$\beta$ ratios on average, followed
by the SDSS quasars and then by the broad \ion{Mg}{2} galaxies.

A
[\ion{O}{3}]/H$\beta$ ratio greater than 0.3 dex has
been used as a 
diagnostic to identify Type 2 quasar candidates from SDSS at $z>0.3$
where the traditional BPT selection is untenable due to [\ion{N}{2}]
and H$\alpha$ being redshifted beyond the SDSS wavelength cutoff \citep{zakamska2003,reyes2008}. We plot this demarcation as a red
dashed line in Figure \ref{o3_hb_hist}  and find that a majority of sources in all
our samples where the narrow lines are significantly detected exceed this threshold, including 61\% of broad \ion{Mg}{2}
galaxies. Compared with Type 1 quasars from SDSS, broad \ion{Mg}{2}
galaxies have similar narrow line properties: $\sim$30-35\% of these
galaxies have [\ion{O}{3}] and narrow H$\beta$ detections and
[\ion{O}{3}]/H$\beta$ ratios indicative of photoionization by an AGN continuum.

\begin{deluxetable*}{lll}
\tablecaption{\label{o3_hb_tab}[\ion{O}{3}]/H$\beta$ Summary}
\tablehead{
  \colhead{Sample} & \colhead{Mean log([\ion{O}{3}]/H$\beta$) }&
  \colhead{Percent of sample where log([\ion{O}{3}]/H$\beta$) $>$ 0.3
    dex} \\
& \colhead{(dex)}}
\startdata
Broad \ion{Mg}{2} Galaxies & $0.40 \pm 0.39$ & 61\%\\
Sample Size & 252 & 153\\
\tableline
SDSS Quasars & $0.55 \pm 0.37$ & 77\% \\
Sample Size & 1702 & 1308 \\
\tableline
Bright-State Changing-Look AGN & $0.79 \pm 0.25$ & 88\% \\
Sample Size & 8 & 7 \\
\tableline
Dim-State Changing-Look AGN & $0.88 \pm 0.25$ & 88\% \\
Sample Size & 8 & 7\\
\enddata
\end{deluxetable*}

\section{Conclusions}
Through a combination of pre-selection and rigourous spectral
decomposition, we identified a sample of 494 broad \ion{Mg}{2}
galaxies from SDSS that are unusual for having broadend (FWHM $>$ 1200 km
s$^{-1}$) \ion{Mg}{2} $\lambda$2798
\angstrom\ emission with weak or absent broad Balmer lines. We
compared the derived spectral properties of these galaxies with
a sample of 3613 Type 1 quasars, a sample of 15 bright-state
changing-look AGN, and a sample of 11 dim-state changing-look AGN selected from SDSS to
test the hypothesis that these broad \ion{Mg}{2} galaxies are fading
AGN, i.e., they are on the tail end of the changing-look AGN sequence
\citep[e.g.,][]{Guo2020a}.

Broad \ion{Mg}{2} galaxies have a
number of properties that are consistent with the paradigm that they
are fading AGN:
\begin{itemize}
  \item Broad \ion{Mg}{2} galaxies have lower AGN luminosities
    (average log ($\lambda L_{\rm 3000}$/erg s$^{-1}$) $\sim$ 43.7) than
    SDSS quasars and changing-look AGN  (average log ($\lambda L_{\rm
      3000}$/erg s$^{-1}$) $\sim$ 44.2-44.4), consistent with the
    picture
    that the AGN continuum is weakening. 
    \item The accretion rate (parameterized by $\lambda_{\rm Edd}$) is
      similar between broad \ion{Mg}{2} galaxies and dim-state
      changing-look AGN (average log($\lambda_{\rm Edd}$) $\sim -2.0$) and is less than the accretion rate of SDSS quasars and bright-state changing
      look AGN  (average log($\lambda_{\rm Edd}$) $\sim -1.5$). These trends align with the prevailing theory that
      changing-look AGN activity, and presence of broad Balmer lines,
      is driven by variability in the accretion energy output
      \citep{KoristaGoad2004,Elitzur2014}.
      \item The average Eddington ratio
      for both broad
      \ion{Mg}{2} galaxies and dim-state changing-look AGN (log($\lambda_{\rm Edd}$) $\sim -2.0$), though with large
        scatter) is the approximate theoretical boundary
      at which the inner accretion disk truncates and is replaced by
      an accretion flow in stellar-mass black holes, which is a model that is invoked as
      a physical driver of changing-look AGN activity \citep{Noda2018,Ruan2019,jana2025,jana2026}.
       \item Similar to typical Type 1 quasars \citep{Dietrich2002}, broad \ion{Mg}{2}
         galaxies display a strong Baldwin effect for \ion{Mg}{2} (Figure
         \ref{mgii_baldwin}) but not for the Balmer lines (Figures
         \ref{hb_baldwin}-\ref{ha_baldwin}, Table \ref{baldwin_effect}),
         consistent with these galaxies being part of the changing-look
         AGN sequence \citep{Guo2020a}.
         \item Using the ratio of [\ion{O}{3}] 5007 \angstrom\ and
           narrow H$\beta$ as a proxy of the ionization field
           hardness, we find that broad \ion{Mg}{2} galaxies have
           ratios consistent with SDSS quasars (Table \ref{o3_hb_tab},
           Figure \ref{o3_hb_hist}), and that most galaxies have ratios
           above 0.3 dex which is a demarcation used to identify Type
           2 quasars from non-active galaxies \citep{zakamska2003,reyes2008}.
         \end{itemize}

 According to the theoretical responsivity of the broad H$\beta$, H$\alpha$, and
 \ion{Mg}{2} lines, as the AGN continuum fades, the broad components of H$\beta
 \rightarrow$ H$\alpha \rightarrow$\ion{Mg}{2} will sequentially
 disappear \citep{KoristaGoad2004}. As a changing-look AGN transitions
 into a dim state and the Eddington ratio stays below a critical value
 of $\sim$1\%, perhaps associated with a radiatively inefficient accretion flow, the broad Balmer lines will disappear yet broad
 \ion{Mg}{2} emission will remain as a relic of previous episodes of
 powerful AGN activity. The observed properties of broad \ion{Mg}{2}
 galaxies are consistent with this changing-look AGN
 interpretation. Such galaxies may
 represent the stage prior to the complete disappearance of broad
 lines in the changing-look sequence, where as the accretion rate
 continues to plunge, the broad \ion{Mg}{2} emission can weaken
 \citep{guo2019,guo2024} and may altogether fade at a low enough
 Eddingtion ratio.  We propose that
study of broad \ion{Mg}{2} galaxies can provide insight into accretion
physics and, when carefully controlling for
selection functions, may probe the AGN duty cycle.

\begin{acknowledgments}
 Funding for the Sloan Digital Sky Survey V has been provided by the Alfred P. Sloan Foundation, the Heising-Simons Foundation, the National Science Foundation, and the Participating Institutions. SDSS acknowledges support and resources from the Center for High-Performance Computing at the University of Utah. SDSS telescopes are located at Apache Point Observatory, funded by the Astrophysical Research Consortium and operated by New Mexico State University, and at Las Campanas Observatory, operated by the Carnegie Institution for Science. The SDSS web site is \url{www.sdss.org}.

SDSS is managed by the Astrophysical Research Consortium for the Participating Institutions of the SDSS Collaboration, including the Carnegie Institution for Science, Chilean National Time Allocation Committee (CNTAC) ratified researchers, Caltech, the Gotham Participation Group, Harvard University, Heidelberg University, The Flatiron Institute, The Johns Hopkins University, L'Ecole polytechnique f\'{e}d\'{e}rale de Lausanne (EPFL), Leibniz-Institut f\"{u}r Astrophysik Potsdam (AIP), Max-Planck-Institut f\"{u}r Astronomie (MPIA Heidelberg), Max-Planck-Institut f\"{u}r Extraterrestrische Physik (MPE), Nanjing University, National Astronomical Observatories of China (NAOC), New Mexico State University, The Ohio State University, Pennsylvania State University, Smithsonian Astrophysical Observatory, Space Telescope Science Institute (STScI), the Stellar Astrophysics Participation Group, Universidad Nacional Aut\'{o}noma de M\'{e}xico, University of Arizona, University of Colorado Boulder, University of Illinois at Urbana-Champaign, University of Toronto, University of Utah, University of Virginia, Yale University, and Yunnan University.  
\end{acknowledgments}

\vspace{5mm}
\facilities{Sloan}

\software{astropy \citep{astropy2013,astropy2018,astropy2022},
  PyQSOFit  \citep{guo18,shen19},
  SciPy \citep{scipy},
  SciServer \citep{Taghizadeh-Popp20}
}

\appendix
We ran \textsc{PyQSOFit} with a range of
Monte Carlo (MC) samples to test the impact on the reported uncertainties of the
fitted parameters. When using MC for the error estimation,
\textsc{PyQSOFit} perturbs the input flux array based on the error
array of the spectrum $N_{\rm samp}$ times, where $N_{\rm samp}$ is the number of
MC samples. \textsc{PyQSOFit} then fits these resampled spectra and
reports the 16th and 84th percentile distribution of the MC samples
as the 1$\sigma$ uncertainty in the output file. However, the reported measurements are from the initial maximum
likelihood estimation (MLE) fit to the data.

For this test, we chose five sources at random from
each of the four samples we considered in this work: broad
\ion{Mg}{2} galaxies, SDSS Type 1 quasars, bright-state changing-look
AGN, and dim-state changing look AGN. We fitted the
spectra for these 20 sources with \textsc{PyQSOFit} with MC samples
set to 50 (as often used in the literature as
cited in the main text and as used in this work), 100, 200, 500, and 1000. We investigated how
the reported uncertainty, measurement precision, agreement between
the measurements and the median of posterior distributions, and compute time
vary as the number of MC samples increases. For this exercise, we focus on
the parameters of the broad \ion{Mg}{2} line used in this work (FWHM, EW,
and flux) and $\lambda L_{\rm 3000}$, though we note that
\textsc{PyQSOFit} only saves the posterior distributions for the
emission line measurements so we are unable to test the level of
agreement between the measured $\lambda L_{\rm 3000}$ value and the
median of the posterior distributions.

From this testing, we find that the reported measurements are not
significantly better constrained when increasing MC samples above 50
to warrant the substantial additional investment in computing time
when increasing $N_{\rm samp}$.

\section{Uncertainty Accuracy}

We explore how the reported uncertainties vary as the number of MC
samples increases. We may expect that since a higher number of MC samples
probes a larger parameter space, the error bars associated with a
higher number of MC samples could be larger than than those with fewer
samples. We thus take the uncertainty associated with $N_{\rm
  samp}$=1000 as our reference uncertainty ($\sigma_{\rm 1000}$) for
this excercise and calculate the normalized uncertainty, or
$\sigma_{\rm s}$/$\sigma_{\rm 1000}$, for each measured parameter,
where $\sigma_{\rm s}$ is the uncertainty reported when running
\textsc{PyQSOFit} with MC samples set to 
$s$ = 50, 100, 200, 500, 1000.

\begin{figure}
  \begin{center}
 \includegraphics[scale=0.85]{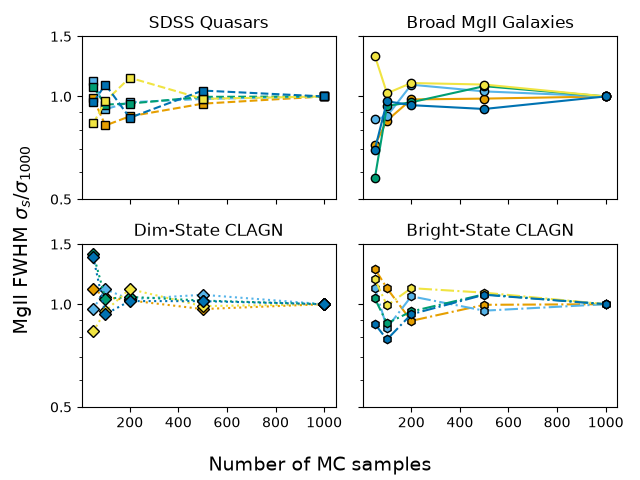}
 \caption{\label{norm_unc_mgii_fwhm} Normalized uncertainty (i.e., $\sigma_{\rm s}$/$\sigma_{\rm 1000}$,
   where $\sigma$ is the reported error, $s$ corresponds
   to the number of MC samples used by \textsc{PyQSOFit} (50, 100,
   200, 500, 1000) and $\sigma_{\rm 1000}$ is the reference
   uncerstainty for N$_{\rm samp}$ = 1000) of the
   broad \ion{Mg}{2} line FWHM. The normalized uncertainty
   is plotted as a function of N$_{\rm samp}$. Five sources were
   chosen from random from each sample (SDSS quasars,top left; broad
   \ion{Mg}{2} galaxies, top right; dim-state changing-look AGN,
   bottom left;  bright-state
   changing-look AGN, bottom right). The normalized uncertainty is
   plotted individually for each source, with different colors for the
   different sources. We find that a lower number of MC samples does
   not necessarily under-estimate the uncertainty and that the
   measured errors across MC samples often agrees with the reference
   value within 10-20\%, though it can be as high as $\sim$40\% for
   the broad \ion{Mg}{2} galaxies.}
\end{center}
\end{figure}

\begin{figure}
  \begin{center}
 \includegraphics[scale=0.85]{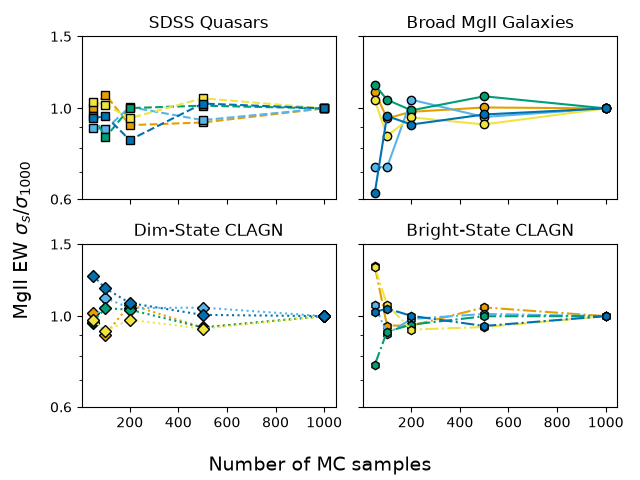}
 \caption{\label{norm_unc_mgii_ew} Same as Figure
   \ref{norm_unc_mgii_fwhm} but for the broad \ion{Mg}{2} line EW.}
\end{center}
\end{figure}

\begin{figure}
  \begin{center}
 \includegraphics[scale=0.85]{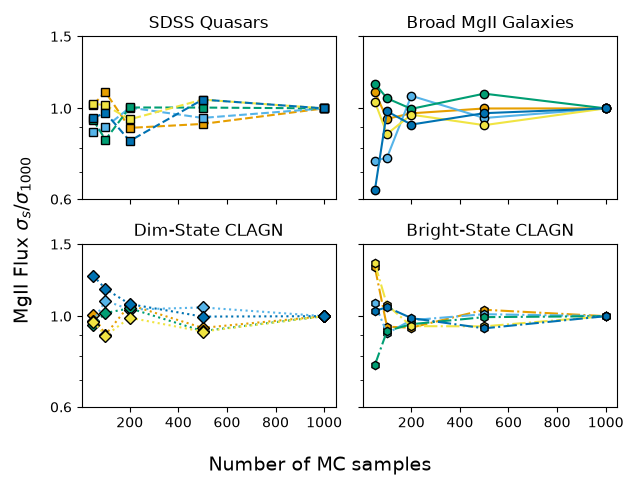}
 \caption{\label{norm_unc_mgii_flux} Same as Figure
   \ref{norm_unc_mgii_flux} but for the broad \ion{Mg}{2} line flux.}
\end{center}
\end{figure}

\begin{figure}
  \begin{center}
 \includegraphics[scale=0.85]{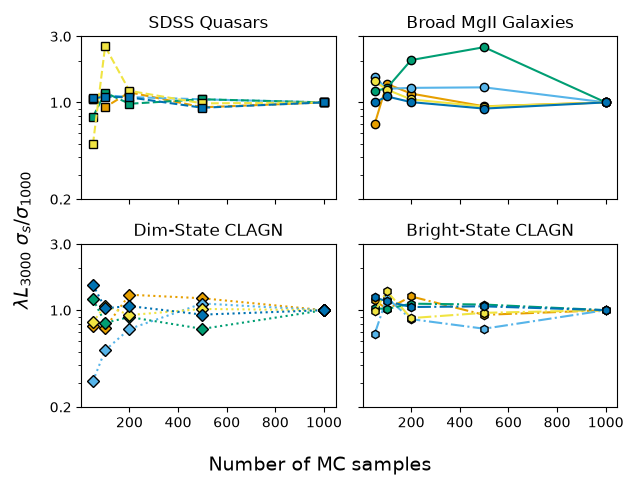}
 \caption{\label{norm_unc_l3000} Same as Figure
   \ref{norm_unc_mgii_flux} but for $\lambda L_{\rm 3000}$. There are
   larger variations in the relative uncertainty across $N_{\rm samp}$
 compared with the broad \ion{Mg}{2} line parameters.}
\end{center}
\end{figure}

We plot the normalized uncertainty of 
the broad \ion{Mg}{2} line FWHM, EW, and flux in Figures \ref{norm_unc_mgii_fwhm} - \ref{norm_unc_mgii_flux}, and
of $\lambda L_{\rm 3000}$ in Figure \ref{norm_unc_l3000}. We have separate panels for
each sample, to search for differences across source type, and we plot
each source separately.

One takeaway is that the normalized uncertainty does not monotonically
increase to the reference value. In fact, in some cases the reported
error when $N_{\rm samp}$=50 exceeds that of the reference value,
indicating that error bars are not necessarily under-estimated
with a lower number of MC samples. We also see that the uncertainties
on the continuum luminosities are less constrained than those on the
line fit parameters, i.e., there is wider variation in the normalized
uncertainty across $N_{\rm samp}$. In many cases, the reported
uncertainties on the \ion{Mg}{2} line fit parameters when $N_{\rm samp}$ = 50 agree with the reference value
within 10-20\%, though this can vary up to 50\%. However, as discussed
below, increasing $N_{\rm samp}$ above 50 comes at a considerable cost
in computing time which is exacerbted when fitting more than a handful
of sources.

\clearpage
\section{Comparison of Measurement Precision across MC Samples}

\begin{figure}
  \begin{center}
 \includegraphics[scale=0.85]{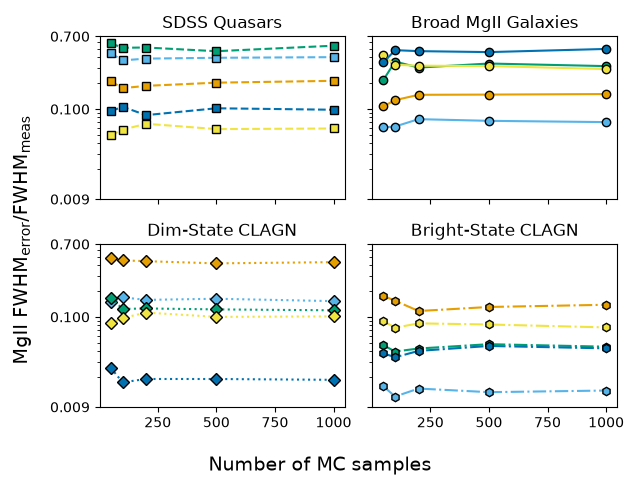}
 \caption{\label{fracerr_mgii_fwhm} Relative uncertainty of the
   broad \ion{Mg}{2} line FWHM, i.e., FWHM$_{\rm error}$/FWHM$_{\rm meas}$
   as a function of $N_{\rm samp}$. There is little change in the
   measurement precision with increasing MC samples, with most
   variations being under 10\%. Color coding same as Figure \ref{norm_unc_mgii_fwhm}.
}
\end{center}
\end{figure}

\begin{figure}
  \begin{center}
 \includegraphics[scale=0.85]{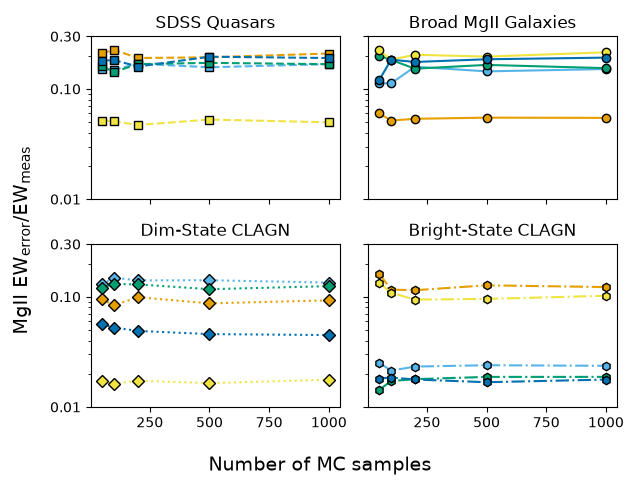}
 \caption{\label{fracerr_mgii_ew} Same as Figure
   \ref{fracerr_mgii_fwhm} but for the broad \ion{Mg}{2} line EW.
}
\end{center}
\end{figure}

\begin{figure}
  \begin{center}
 \includegraphics[scale=0.85]{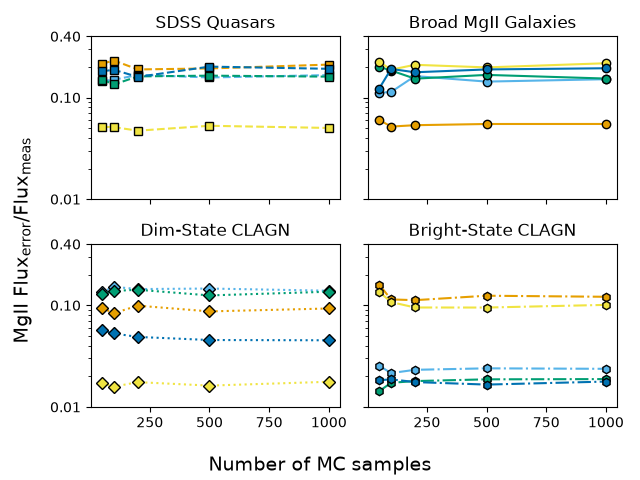}
 \caption{\label{fracerr_mgii_flux} Same as Figure
   \ref{fracerr_mgii_fwhm} but for the broad \ion{Mg}{2} line flux.
}
\end{center}
\end{figure}

\begin{figure}
  \begin{center}
 \includegraphics[scale=0.85]{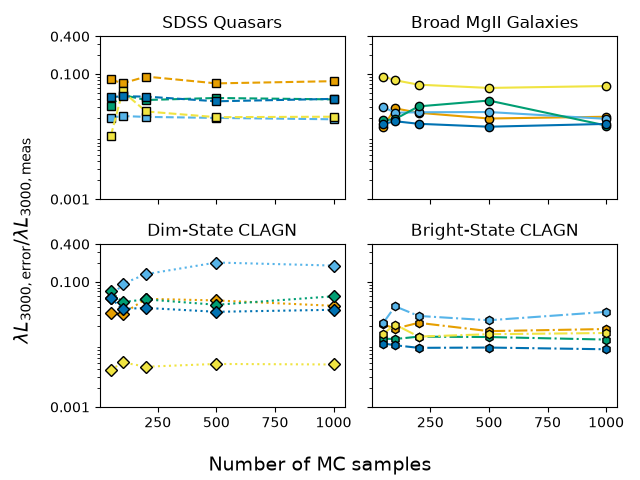}
 \caption{\label{fracerr_l3000} Same as Figure
   \ref{fracerr_mgii_fwhm} but for $\lambda L_{\rm 3000}$.
}
\end{center}
\end{figure}

We investigated how the precision of a reported \textsc{PyQSOFit}
measurement varies when increasing the number of MC samples. We use
the relative error, or the uncertainty normalized by the measurement,
for this exercise, and plot it against the number of MC samples for
the broad \ion{Mg}{2} FWHM, EW, and flux in Figures
\ref{fracerr_mgii_fwhm} - \ref{fracerr_mgii_flux} and $\lambda
L_{\rm 3000}$ in Figure \ref{fracerr_l3000}.

As demonstrated in Figures \ref{fracerr_mgii_fwhm} - \ref{fracerr_l3000}, there is generally little change in
the measurement precision as the number of MC samples
increases, varying by under 10\% for most sources. This exercise indicates
that increasing the number of MC samples above 50 does not
necessarily result in measurements that are better constrained.

\section{Comparison of Measurements with the Median of the Posterior
  Distribution}
\textsc{PyQSOFit} offers the capability to access the line fitting results from each
MC resampling run. We were thus able to calculate the median of the
posterior distributions of the broad \ion{Mg}{2} line fit parameters
for each \textsc{PyQSOFit} run where $N_{\rm samp}$ = 50, 100, 200, 500,
1000.

\begin{figure}
  \begin{center}
 \includegraphics[scale=0.85]{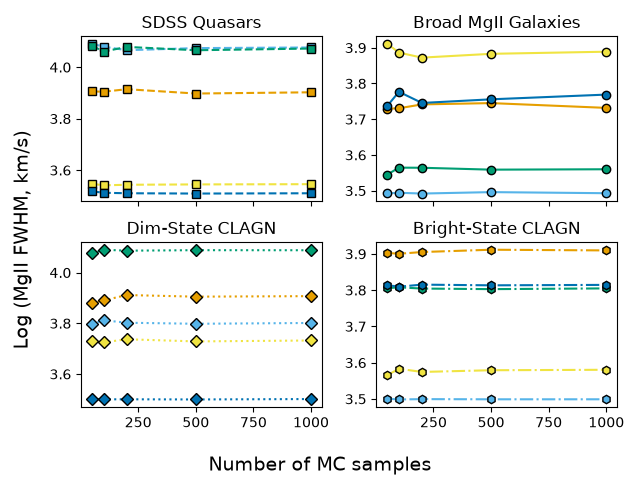}
\caption{\label{post_mgii_fwhm} Median of the posterior distribution
  of the fitted broad \ion{Mg}{2} FWHM for $N_{\rm samp}$ = 50, 100,
  200, 500, 1000. We find that the median FWHM does not vary with
  $N_{\rm samp}$. Color coding same as Figure \ref{norm_unc_mgii_fwhm}.}
\end{center}
\end{figure}

\begin{figure}
  \begin{center}
 \includegraphics[scale=0.85]{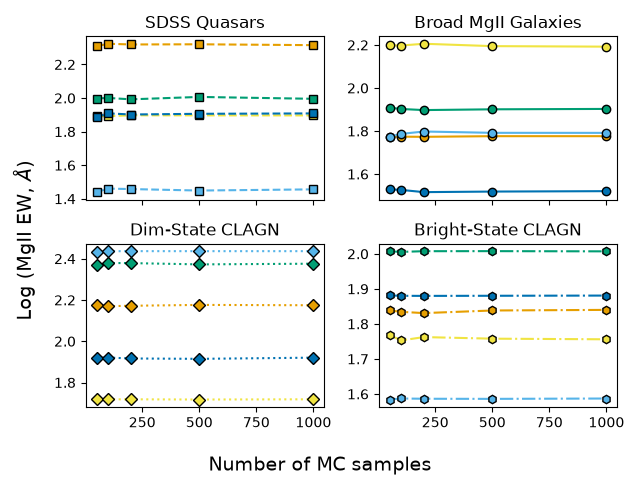}
\caption{\label{post_mgii_ew} Same as Figure \ref{post_mgii_fwhm} but for the fitted
  broad \ion{Mg}{2} EW.}
\end{center}
\end{figure}

\begin{figure}
  \begin{center}
 \includegraphics[scale=0.85]{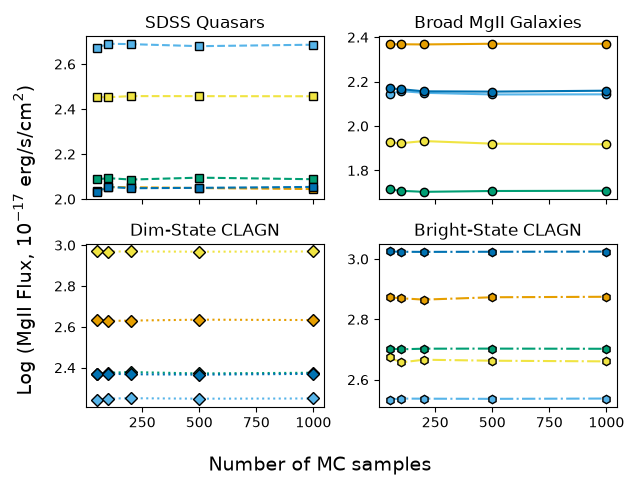}
\caption{\label{post_mgii_flux} Same as Figure \ref{post_mgii_fwhm} but for the fitted
  broad \ion{Mg}{2} flux.}
\end{center}
\end{figure}

In Figures \ref{post_mgii_fwhm} - \ref{post_mgii_flux}, we plot the median of the posterior distributions
of the broad \ion{Mg}{2} FWHM, EW, and flux, respectively, as a
function of $N_{\rm samp}$. We find that these values remain
relatively constant with $N_{\rm samp}$, indicating that increasing
the number of MC samples does not appreciably shift the 
parameter space from which uncertainties are calculated and reported.

\begin{figure}
  \begin{center}
 \includegraphics[scale=0.85]{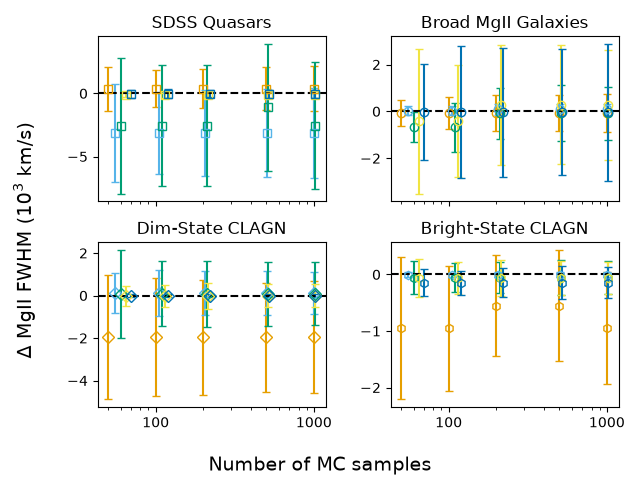}
 \caption{\label{resid_post_mgii_fwhm} Difference between the measured
  broad \ion{Mg}{2} FWHM and the median of the posterior distribution ($\Delta$ \ion{Mg}{2} FWHM = \ion{Mg}{2} FWHM$_{\rm
  measured}$ - \ion{Mg}{2} FWHM$_{\rm median}$) for $N_{\rm samp}$ = 50, 100,
  200, 500, 1000. We plot this difference with a slight offset in the
  $x$-axis for each source to ease visualization. Though 1-2 sources
  for most samples show a sysetmatic offset between the median value
  and the median of the posterior distribution, increasing $N_{\rm
    samp}$ does not decrease this offset. All measurements are
  consistent with the median of the posterior distribution within the
  reported uncertainties for all MC samples. Increasing $N_{\rm samp}$
  above 50 does not correspond to more accurate measurements. The
  color coding is same as Figure \ref{norm_unc_mgii_fwhm}.}
\end{center}
\end{figure}

\begin{figure}
  \begin{center}
 \includegraphics[scale=0.85]{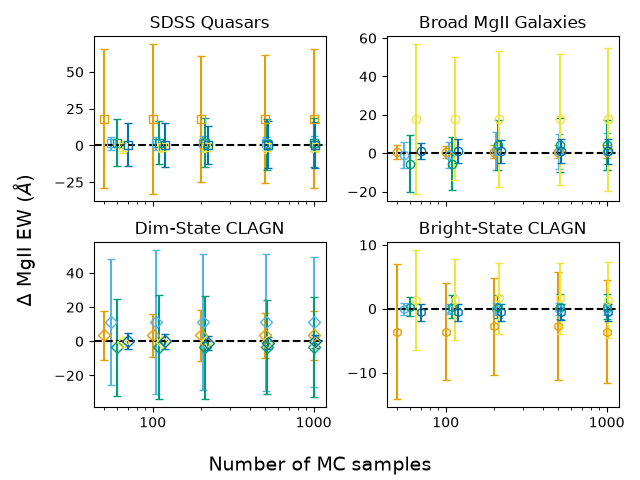}
 \caption{\label{resid_post_mgii_ew} Same as Figure
   \ref{resid_post_mgii_fwhm} but for the broad \ion{Mg}{2} EW. }
\end{center}
\end{figure}

\begin{figure}
  \begin{center}
 \includegraphics[scale=0.85]{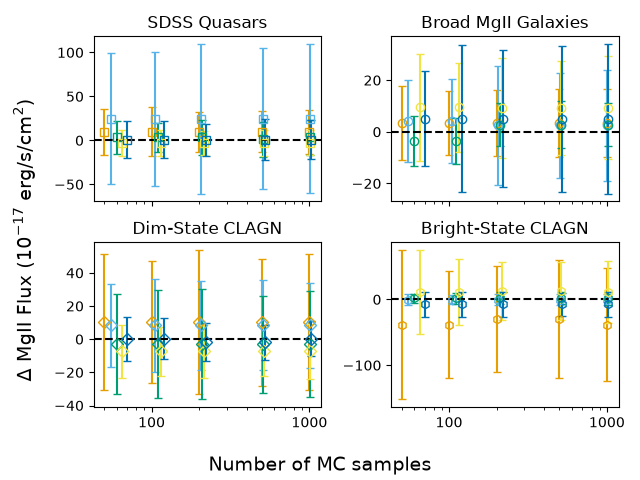}
 \caption{\label{resid_post_mgii_flux} Same as Figure
   \ref{resid_post_mgii_fwhm} but for the broad \ion{Mg}{2} flux. }
 \end{center}
\end{figure}

But how well do the reported measurements (from the initial MLE fit to
the spectrum) agree with the median of the
posterior distribution given the reported uncertainties? To
answer this question, we select the median of the posterior
distribution for $N_{\rm samp}$ = 1000 as a reference value and
calculate the the difference between the measured value and this
reference (e.g., $\Delta$ \ion{Mg}{2} FWHM = \ion{Mg}{2} FWHM$_{\rm
  measured}$ - \ion{Mg}{2} FWHM$_{\rm median}$). We plot these
residuals and associated measurement uncertainties as a function of
$N_{\rm samp}$ in Figures \ref{resid_post_mgii_fwhm} -
\ref{resid_post_mgii_flux}. We find that some sources have a systematic
offset between the measured value and the median of the posterior
distribution, but when this offset does occur, it
persists amongst all MC samples. And in all cases, the difference
between the measurement and the median of the posterior distribution
is consistent with 0 within the reported uncertainties. We thereby
conclude that increasing $N_{\rm samp}$ above 50 does not necessarily improve
measurement accuracy.

\section{Compute time}

Increasing the number of MC samples in a \textsc{PyQSOFit} run incurs
a cost in computing time: each additional sample requires resampling
the spectrum, fitting the resampled spectrum, and storing the
results. We recorded the amount of computing time for each
\textsc{PyQSOFit} run in this exercise.

\begin{figure}
  \begin{center}
 \includegraphics[scale=0.85]{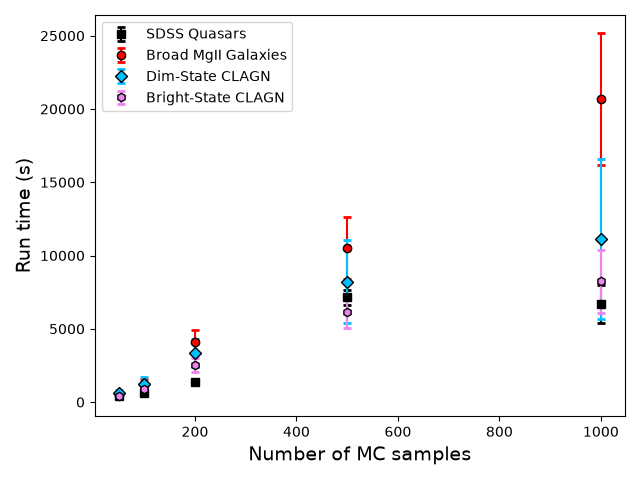}
 \caption{\label{mc_run_times} Average \textsc{PyQSOFit} run times for
 sources in the SDSS quasar sample (black squares), broad \ion{Mg}{2}
 galaxy sample (red circles), dim-state changing-look AGN (blue
 diamonds), and bright-state changing-look AGN (purple hexagons) as a
 function of the number of MC samples used for the error
 estimation. The host-galaxy dominated spectra (broad \ion{Mg}{2}
 galaxies, dim-state changing-look AGN) take a longer time to fit than
AGN-dominated spectra. The average running time per source across all samples ranges from $\sim$8 minutes ($N_{\rm samp}$ = 50) to over 3 hours ($N_{\rm samp}$ =
1000). The increase cost in computing time is significant and not
commensurate with improvements in error estimation or measurement
precision or accuracy.}
 \end{center}
\end{figure}

In Figure \ref{mc_run_times}, we calculated
the average compute time for all sources in each sample, where the
error bars represent the standard deviation. The average time to run
\textsc{PyQSOFit} is higher for galaxy-dominated spectra (i.e., broad
\ion{Mg}{2} galaxies, followed by dim-state changing-look AGN) than
for AGN-dominated spectra. We find that the average time to fit
spectra with \textsc{PyQSOFit} ranges from $\sim$6-10 minutes per
source for $N_{\rm samp}$ = 50, $\sim$10-20 minutes per source for
$N_{\rm samp}$ = 100, $\sim$20-70 minutes per source for $N_{\rm
  samp}$ = 200, $\sim$1.7 - 2.9 hours per source for $N_{\rm samp}$ =
500, and $\sim$1.9 - 5.7 hours per source for $N_{\rm samp}$ = 1000.

We conclude that increasing the number of MC samples above 50 incurs a
signifcant cost in computing time that is not associated with
demonstrable improvements in fit parameter uncertainties or
measurement precision or accuracy.

\end{document}